\documentclass[%
superscriptaddress,
 amsmath,amssymb,
 aps,
 pra,
twocolumn
]{revtex4-2}

\usepackage{graphicx}%
\usepackage{dcolumn}%
\usepackage{bm}%
\usepackage{hyperref}%
\hypersetup{
	colorlinks,
	linkcolor={blue},
	citecolor={blue},
	urlcolor={blue}
}
\usepackage{mathtools}
\usepackage{cleveref}
\usepackage[caption=false]{subfig}
\usepackage{braket}

\newcommand{\twodots}{\mathinner {\ldotp \ldotp}}

\usepackage{algorithm, algorithmic}

\usepackage{xcolor}

\begin{document}

  \title{Improved decoding of circuit noise and fragile boundaries of tailored surface codes}
        
\date{\today}

\author{Oscar Higgott}
\email{oscar.higgott.18@ucl.ac.uk}
\affiliation{
  Department of Physics and Astronomy, University College London, Gower Street, London WC1E 6BT, United Kingdom
 }
 \affiliation{AWS Center for Quantum Computing, Cambridge CB1 2GA, United Kingdom}

 \author{Thomas C. Bohdanowicz}
 \affiliation{Goldman, Sachs \& Co., New York, NY, USA}
 \affiliation{AWS Center for Quantum Computing, Pasadena, CA, USA}

\author{Aleksander~Kubica}
\affiliation{AWS Center for Quantum Computing, Pasadena, CA, USA}
\affiliation{California Institute of Techonology, Pasadena, CA, USA}

\author{Steven T. Flammia}
\affiliation{AWS Center for Quantum Computing, Pasadena, CA, USA}
\affiliation{California Institute of Techonology, Pasadena, CA, USA}

\author{Earl T.~Campbell}
\affiliation{AWS Center for Quantum Computing, Cambridge CB1 2GA, United Kingdom}
\affiliation{Riverlane, Cambridge, United Kingdom}
\affiliation{Department of Physics and Astronomy, University of Sheffield, Sheffield S3 7RH, United Kingdom}

\begin{abstract}
Realizing the full potential of quantum computation requires quantum error correction (QEC), with most recent breakthrough demonstrations of QEC using the surface code. QEC codes use multiple noisy physical qubits to encode information in fewer logical qubits, enabling the identification of errors through a decoding process. This process increases the logical fidelity (or accuracy) making the computation more reliable. However, most fast (efficient runtime) decoders neglect important noise characteristics, thereby reducing their accuracy. In this work, we introduce decoders that are both fast and accurate, and can be used with a wide class of QEC codes including the surface code.  Our decoders, named belief-matching and belief-find, exploit all noise information and thereby unlock higher accuracy demonstrations of QEC. Using the surface code threshold as a performance metric, we observe a threshold at 0.94\% error probability for our decoders, outperforming the 0.82\% threshold for a standard minimum-weight perfect matching decoder. 
We also tested our belief-matching decoders in a theoretical case study of codes tailored to a biased noise model. We find that the decoders led to a much higher threshold and lower qubit overhead in the tailored surface code with respect to the standard, square surface code. Surprisingly, in the well-below threshold regime, the rectangular surface code becomes more resource-efficient than the tailored surface code, due to a previously unnoticed phenomenon that we call ‘fragile boundaries’. Our decoders outperform all other fast decoders in terms of threshold and accuracy, enabling better results in current quantum error correction experiments and opening up new areas for theoretical case studies. 
\end{abstract}

\maketitle

\section{Introduction}

Quantum error correction (QEC) is an essential ingredient for building a useful quantum computer. Using QEC we can exponentially reduce the probability of a computational failure to any desired level by increasing the number of qubits used.  We can use QEC whenever the probability of failure ($p$) for each quantum logic gate is below some value known as the ``threshold" ($p_{\mathrm{th}}$). The most widely studied QEC code is the surface code, which has a high threshold and uses gates performed between nearest neighbour qubits arranged in a two dimensional grid~\cite{dennis2002topological,kitaev2003fault}.
Consequently, the surface code is particularly amenable to experimental implementations, as highlighted by recent demonstrations \cite{krinner2022realizing,google2023suppressing}.

QEC codes require \textit{decoders}, which are algorithms running on a classical computer that determine where errors occurred.  The accuracy of a decoder quantifies how good it is at correctly determining where errors occurred. 
A more accurate decoder can increase the value of the threshold for a QEC code, as well as reducing the number of physical qubits required to achieve a desired logical fidelity below threshold.
Improving the accuracy of decoders can therefore lead to less demanding hardware requirements.
Speed is also an important decoder metric. Ideally, a decoder will have an expected running time that scales linearly or almost-linearly with the size of the problem, since the decoder must keep up with the quantum hardware to prevent an exponentially growing decoding backlog~\cite{terhal2015quantum,skoric2022parallel}. We informally call these fast decoders.  Previous decoders have either been highly accurate~\cite{bravyi2014efficient,bacon2017sparse,thomthesis} or fast ~\cite{dennis2002topological,fowler2013minimum,delfosse2021almost,huang2020fault,higgott2023sparse,wu2023fusion,liyanage2023scalable} but not both. Here, we propose decoders that are both fast and accurate.

\begin{figure*}
    \centering
    \includegraphics{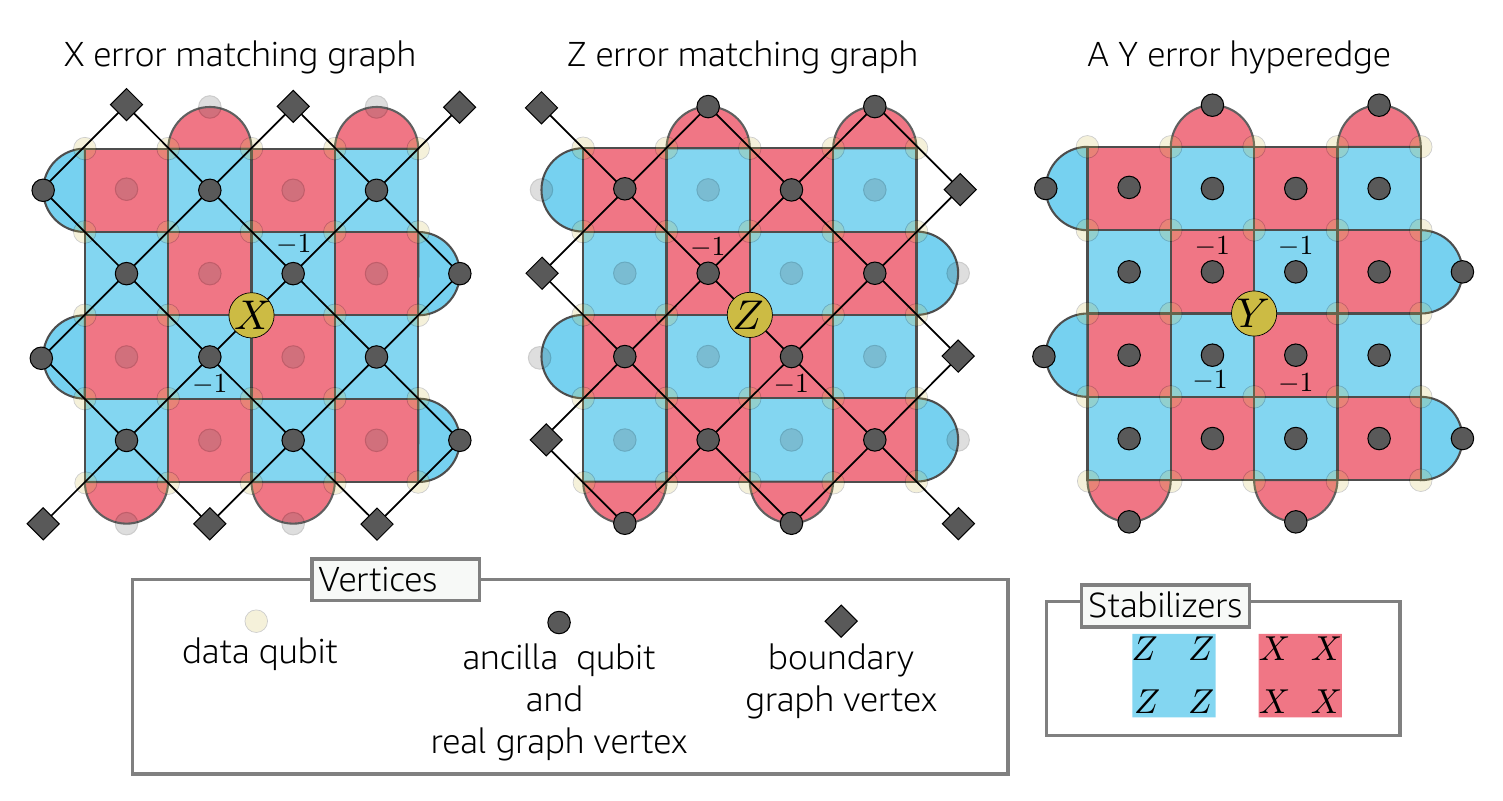}
    \caption{The MWPM decoding problem for a distance 5 surface code. Left: the $X$ error matching graph, where we associate a node with each stabilizer and an edge $(u, v)$ with each $X$ error, where $u$ and $v$ are the stabilizers that the error anti-commutes with.
    If an $x$ error anti-commutes with a single stabilizer $u$ we represent it with an edge $(u,b)$ between $u$ and a \textit{boundary} node (each a square node in the diagram).
    Middle: the $Z$ error matching graph is defined similarly, but with an edge for each $Z$ error. Right: A $Y$ error anti-commutes with four stabilizers, so would need to be represented by a hyperedge, and induces correlations between the $X$ and $Z$ matching graphs.}
    \label{fig:fig1}
\end{figure*}

Fast decoders for the surface code, including minimum-weight perfect matching (MWPM)~\cite{dennis2002topological,fowler2013minimum,higgott2023sparse,wu2023fusion} and union-find (UF)~\cite{delfosse2021almost,huang2020fault}, use a coarse approximation of the noise model, ignoring important error mechanisms that are ubiquitous in experiments. For example, both UF and MWPM ignore the possibility of $Y$ errors that introduce correlations between the $X$ and $Z$ decoding problems.
As shown in \Cref{fig:fig1}, either an $X$ or $Z$ error leads to at most a pair of error-detection events, enabling them to be interpreted as edges in a graph called the matching graph. 
In contrast, $Y$ errors lead to 4 error-detection events (\Cref{fig:fig1}) and can not be represented in a matching graph. 
For this reason, the matching graph is only an approximation of the full error model, and as a result MWPM and UF do not have very high accuracy compared to some other (slow) decoders \cite{bravyi2014efficient,bacon2017sparse,thomthesis}.  On the other hand, previously proposed decoders that have high accuracy are slow (with exponentially scaling running time) and already impractical for modest size QEC codes.

In this work, we introduce fast (computationally efficient) and accurate decoders applicable to surface codes. An important subroutine of our decoders is the belief-propagation (BP) algorithm that updates prior beliefs about where errors are most likely to have occurred via an easily parallelisable message passing protocol.  Crucially, our use of BP enables us to exploit all the information present in circuit-level noise models more effectively, handling correlations between the $X$ and $Z$ decoding problems, and thereby achieving higher accuracy than MWPM or UF.  While BP is powerful at exploiting the full noise information, by itself BP often fails to converge to a valid solution.  We show that by marrying belief-propagation (BP) with MWPM or weighted UF we both ensure convergence and make full use of all noise information, thereby boosting accuracy. 

More precisely, whenever BP fails to converge, we use the updated beliefs output by BP to determine the edge weights in a matching graph.
We then decode this re-weighted matching graph either: using MWPM, in which case we refer to the overall decoder as \textit{belief-matching}; or instead using weighted union-find, in which case we name the decoder \textit{belief-find}. Belief-matching has conceptual similarities to the decoder proposed by Criger and Ashraf~\cite{criger2018multi}, which considered a toy noise model with perfect measurement results. 
A key difference of our approach is applicability to real experimental data and circuit-level noise simulations of experiments. We show that belief-matching and belief-find are the most accurate of all known computationally efficient decoders, and belief-find even has an almost-linear (worst-case) running time.  
Our numerical simulations show that the high accuracy of our decoders leads to an increase in the surface code threshold with circuit-level noise from 0.82\% (for MWPM) to 0.94\% (for belief-matching and belief-find). After our work was posted as a preprint, the high accuracy of our decoder for real devices was confirmed by the Google team. In their recent landmark QEC experiment showing logical error suppression~\cite{google2023suppressing}, the Google team tested many decoders and our belief-matching decoder was the only efficient decoder that was accurate enough to observe the desired logical error suppression effect.

\begin{figure*}
    \centering
    \includegraphics{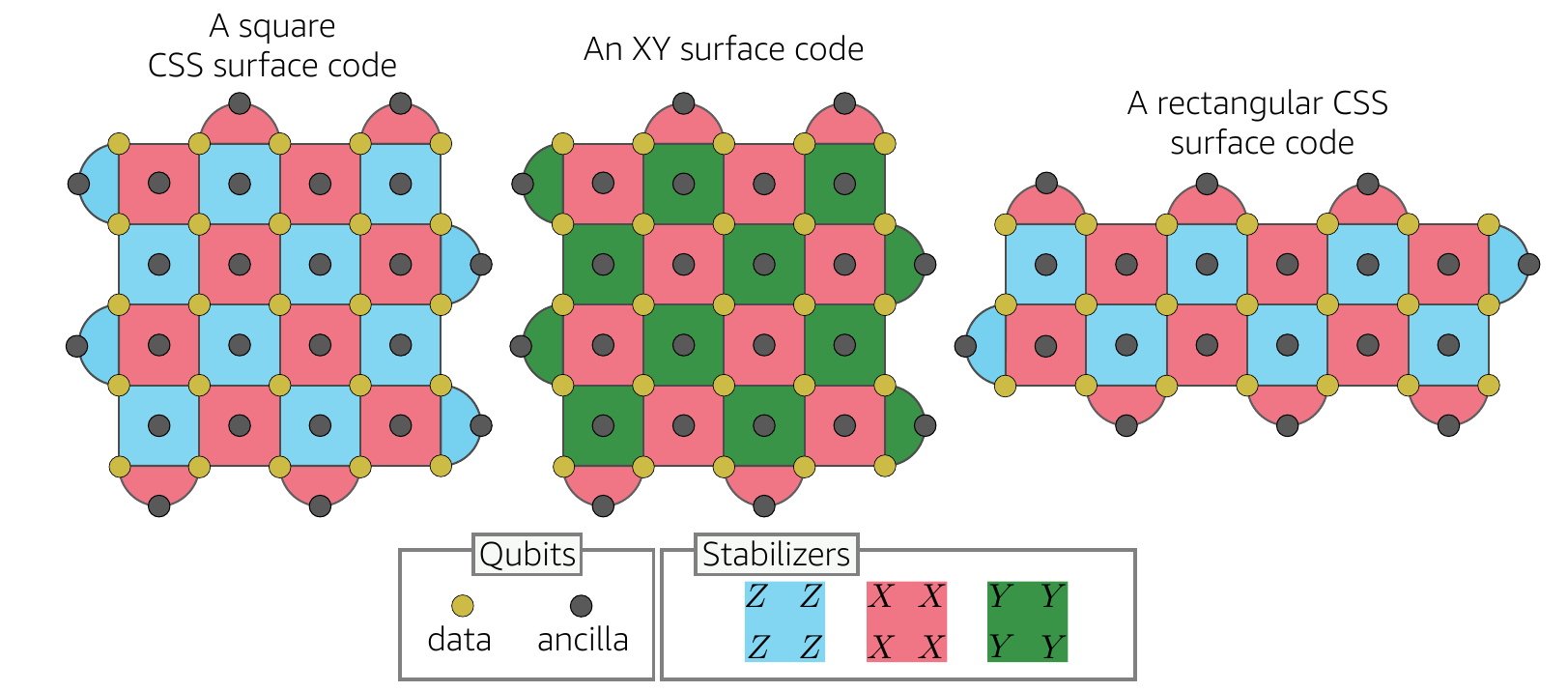}
    \caption{The three code families compared in this work.
    The square, CSS surface code is the most commonly encountered surface code, without any tuning for the noise bias.
    We denote the lattice size by $L$, and here the square CSS and XY surface codes both have $L=5$.
    }
    \label{fig:Codes}
\end{figure*}

Our decoders can be directly used for any QEC code for which the MWPM decoder is applicable, which includes the standard and XY (or tailored) surface codes~\cite{dennis2002topological,tuckett2018ultrahigh} as well as other two-dimensional subspace and subsystem codes~\cite{bacon2006operator,bravyi2012subsystem,breuckmann2016constructions,li20192d,chamberland2020topological,bonilla2020xzzx,higgott2021subsystem, hastings2021dynamically,dua2022clifford,Xu2023,delfosse2014decoding2,kubica2023efficient}. As a case study, we use our decoders to tackle the open problem of determining the optimal variant of the surface code in the presence of biased noise that favours phase errors over bit-flips~\cite{aliferis2008fault, puri2020bias, guillaud2019repetition, chamberland2020building}.  Since other fast decoders do not fully exploit the information in a biased noise setting, this case study illustrates new avenues of research opened by our decoders.  We assess these codes by optimizing for the fewest physical resources required to achieve a desired logical failure rate.
We assume qubits are constrained to a square lattice geometry with boundaries. Several methods have been proposed for exploiting this noise bias information through modifications to the choice of code (see \Cref{fig:Codes}), with the aim of increasing thresholds or reducing the qubit overhead below threshold~\cite{tuckett2018ultrahigh, tuckett2019tailoring, tuckett2020fault, bonilla2020xzzx, darmawan2021practical, higgott2021subsystem}. The codes we consider are square and rectangular surface codes~\cite{dennis2002topological} (referred to as CSS) and a modified surface code for which $Z$ stabilizers are replaced with $Y$ stabilizers~\cite{tuckett2018ultrahigh} that is called the XY surface code.  No previous work has performed a fair comparison of these code families. 

In our case study, we use our new belief-matching algorithm to decode biased circuit-level noise in the XY surface code.  We find that it significantly outperforms MWPM alone, and we observe a threshold of 0.841(6)\% CNOT infidelity for biased circuit-level noise.
This constitutes a $1.69\times$ relative improvement on the 0.498(2)\% threshold observed using MWPM.
Unfortunately, we discovered that the high tolerance of the XY surface code to $Z$ errors is extremely fragile.
This fragility occurs wherever the space-time picture of the XY surface code has a boundary. Consequently, for CNOT infidelities below around 0.4\%, we find the surprising result that \textit{rectangular} CSS surface codes outperform the XY surface codes, owing to the reduction in qubit overhead achieved by optimizing the aspect ratio of the lattice for the CSS surface code.

At the spatial boundary of an XY surface code, we find failure mechanisms that require only $O(\sqrt{n})$ $Z$ errors and a single $X$ or $Y$ error.
We refer to these as fragile spatial boundary errors. Using belief-matching, we present numerical results consistent with the conclusion that these failure mechanisms dominate at lower error rates and finite bias.  The temporal boundaries correspond to logical state preparation (the earliest time boundary) and logical measurement (the latest time boundary).  We also find string-like $Z$ errors that can occur on these temporal boundaries even at infinite bias.  These occur because during logical measurement, we only measure half the code stabilizers (just $X$ type or just $Y$ type) which reduces the protection from errors. Logical state preparation is the mirror image of logical measurement and similarly susceptible to such failure mechanisms.  Temporal boundaries also arise during lattice surgery~\cite{horsman2012surface,litinski2018lattice,litinski2019game,chamberland2021universal} and so these operations are also vulnerable. We refer to this family of errors as fragile temporal boundary errors. None of the prior art reviewed above ~\cite{tuckett2018ultrahigh,tuckett2020fault} considered: the below-threshold error scaling at finite-bias with open boundary conditions; or the error scaling of logical state preparation and measurement errors.  Consequently, our case study is the first to observe the dominant error mechanisms reported here, providing a new insight in how to best design QEC codes for biased noise.

The structure of our paper is as follows.
We introduce some relevant background theory and notation in \Cref{sec:preliminaries}.
In \Cref{sec:decoding_circuit_level} we introduce our belief-matching and belief-find decoders for circuit-level noise and also review the tensor network approach to maximum likelihood decoding.  \Cref{sec:decoding_circuit_level} also presents numerical results for circuit-level depolarising noise.
In \Cref{sec:tailoring_codes} we present our case study results, explaining how fragile boundary errors inhibit the performance of the XY surface code. Finally, in \Cref{sec:conclusion}, we conclude by summarising our key findings and discussing possible future work.

\section{Preliminaries}
\label{sec:preliminaries}

The Pauli group $\mathcal{P}_n$ is the set of all $n$-qubit Pauli operators $P=\alpha P_1\otimes \dots \otimes P_n$, where $\alpha\in\{\pm 1, \pm i\}$ and $P_i\in\{I,X,Y,Z\}$.
The \textit{weight} of a Pauli operator is the number of qubits that it acts on non-trivially.
Here, $X,Y,Z$ are Pauli operators and $I$ is the identity.
A very broad family of quantum error correcting codes are \textit{stabilizer codes}.
A stabilizer code is defined as the joint $+1$-eigenspace of a stabilizer group $\mathcal{S}$, which is an abelian subgroup of $\mathcal{P}_n$ that does not contain $-I$~\cite{gottesman1997stabilizer}.
We can define a stabilizer code using a set of independent generators of its stabilizer group $\mathcal{S}=\langle g_1,g_2,\ldots,g_r \rangle$.
A stabilizer code is a Calderbank-Shor-Steane (CSS) code if its stabilizer group admits a set of generators $g_1, g_2, \ldots, g_r$ such that each generator is either $X$-type or $Z$-type, $g_i\in\{I,X\}^{\otimes n}\cup \{I,Z\}^{\otimes n}$~\cite{calderbank1996good,steane1996error}.
The centralizer $C(\mathcal{S})$ of a stabilizer code is the set of Pauli operators that commute with every element of $\mathcal{S}$.
The elements of $C(\mathcal{S})\setminus \mathcal{S}$ are undetectable logical errors, and the \textit{distance} of a stabilizer code is the minimum weight of any element of $C(\mathcal{S})\setminus \mathcal{S}$. We use the phrase $Z$ distance when considering only $Z$ errors, and similarly for $X$ distance. The generators of the CSS surface code are shown in \Cref{fig:Codes}.
The CSS surface code is particularly amenable to being realised experimentally, in part because its stabilizer generators are low weight (at most four) and geometrically local on a 2D Euclidean surface, such as a quantum computer chip.

Suppose a Pauli error $E\in\mathcal{P}_n$ occurs, which must either commute or anti-commute with a given generator $g_i$ of $\mathcal{S}$.
Measuring each stabilizer generator we obtain a \textit{syndrome} $\sigma(E)$, which is a list of the measured eigenvalues of the generators of $\mathcal{S}$ (each $g_i$ has eigenvalue $-1$ or $1$).
A generator $g_i$ will then measure 1 if it commutes with $E$ and measure -1 if it anti-commutes.
Given the syndrome and a known noise model, a \textit{decoder} makes a prediction $C\in\mathcal{P}_n$ of which error occurred.
If $EC\in\mathcal{S}$ then the decoder has succeeded in correcting the error, whereas if $EC\notin\mathcal{S}$ then a logical error has occurred.
See \Cref{fig:fig1} for examples of some single qubit errors in the surface code.

In practice, the stabilizer generators of the code are measured using a syndrome extraction circuit, and the gates and measurements in this circuit can themselves be faulty.  Allowing for errors to occur anywhere in a syndrome extraction circuit is called circuit-level noise.
To handle faulty measurements in the surface code we repeat each cycle of stabilizer measurements $O(d)$ times (where here $d$ is the code distance) to ensure the stabilizer outcomes can be inferred reliably~\cite{dennis2002topological}.
The syndrome input to the decoder is now determined from \textit{detector} measurements.
A detector is defined to be a linear combination of measurement outcomes in a circuit that would have a deterministic outcome if no noise was present~\cite{gidney2021stim}.
A detector is also referred to as an \textit{error-sensitive event} in the literature~\cite{chen2021calibrated}.
In this example of the surface code, the linear combination of each pair of consecutive stabilizer ancilla measurements is taken to define a detector.
We say that a detector has \textit{flipped} if its binary value differs from the value it would take in an error-free syndrome extraction circuit, and the \textit{syndrome} $\sigma$ is the set detectors that have flipped.

\section{Decoding circuit-level noise}
\label{sec:decoding_circuit_level}
 
Conventional decoders for the CSS surface code treat $X$-type and $Z$-type errors as two independent decoding problems.
If we first assume perfect syndrome measurements, we note that $X$ and $Z$ errors each anti-commute with two stabilizers in the bulk, and can be represented as edges in a matching graph, where the nodes correspond to stabilizer measurements (see \Cref{fig:fig1})~\cite{dennis2002topological,fowler2013minimum}.
We refer to error mechanisms that flip one or two stabilizers (or detectors) as ``graphlike'' since a pair of detectors can be associated with an edge in a graph.
This matching graph, along with the syndrome, can be used to decode efficiently using a MWPM decoder (which finds the most probable physical graphlike error) or union-find decoder (an approximation of MWPM with improved worst-case running time)~\cite{dennis2002topological,fowler2013minimum,higgott2021pymatching,delfosse2021almost,huang2020fault,wu2022interpretation}.
On the other hand, $Y$ errors anti-commute with four stabilizers in the bulk. Therefore, there is no edge (no pair of detectors) corresponding to $Y$ errors, and we instead represent these by a \textit{hyperedge}.  In graph theory, a hyperedge connects more than a pair of vertices. 
However, hyperedges are not supported by MWPM or UF, and so the correlations that $Y$ errors induce between the $X$ and $Z$ decoding problems are not exploited, leading to performance that is far from optimal.  These observations also carry over to the setting of decoding circuit-level noise occurring during syndrome extraction circuits, for which the matching graph is three-dimensional (with time being the third dimension).

Several different approaches have been proposed for handling hyperedge error mechanisms more effectively than MWPM or UF~\cite{fowler2013optimal,bravyi2014efficient,sundaresan2022matching,delfosse2014decoding,criger2018multi,baireuther2018machine,torlai2017neural,meinerz2022scalable,tuckett2020fault, benhemou2023minimising}.
In Ref.~\cite{bravyi2014efficient}, a tensor network decoder was introduced that approximates maximum-likelihood decoding for surface codes.
However, this approach has high computational complexity and assumes error-free syndrome extraction circuits.
\citeauthor{tuckett2020fault} developed a decoder for the surface code tailored to the case where hyperedge error mechanisms dominate over graphlike error mechanisms, finding improved thresholds in the XY surface code at finite and infinite $Z$ bias relative to the MWPM decoder~\cite{tuckett2020fault}.
However, while the performance of the decoder is promising for phenomenological noise, it is not clear how well suited it is to other noise models, such as depolarising noise or general circuit-level errors in syndrome extraction circuits. In Ref.~\cite{criger2018multi} BP was used, along with multi-path summation, to choose edge weights for a MWPM decoder, finding a threshold of 17.76\% for the surface code with depolarising noise and perfect syndrome measurements.  However, Ref.~\cite{criger2018multi} did not consider how to generalise the method to handle noisy gates in the syndrome extraction circuit.

\begin{figure*}
    \centering
    \includegraphics[width=1.8 \columnwidth]{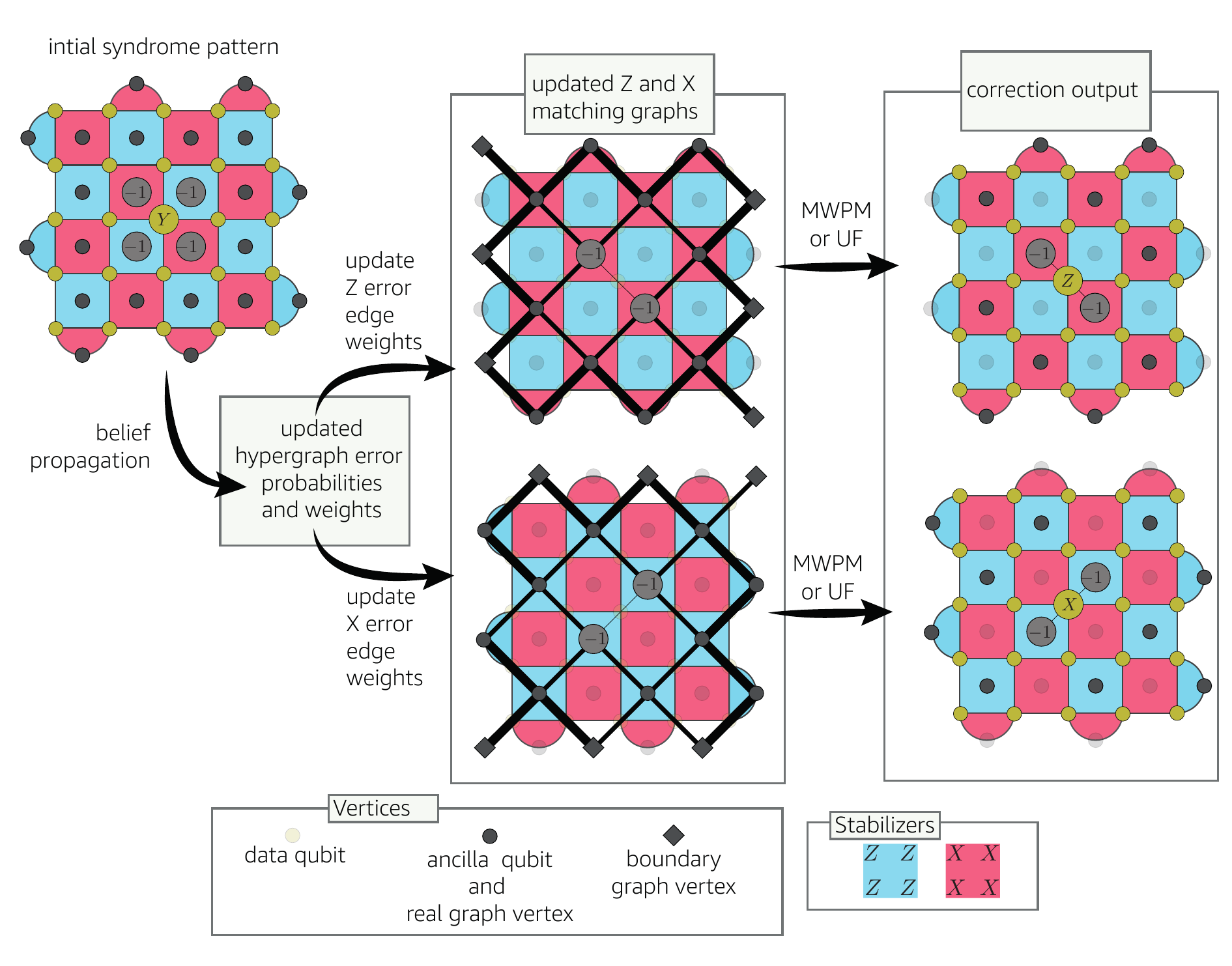}
    \caption{Illustration of belief-matching and belief-find. Given an observed syndrome and an error model, belief propagation is used to estimate the marginal probability that each error mechanism occurred. These updated error probabilities are used to set edge weights (here, thicker edges correspond to higher edge weights) in the $X$ and $Z$ matching graphs, which are then decoded with MWPM (for belief-matching) or weighted UF (for belief-find). In this figure we consider the decoding problem for perfect syndrome measurements for simplicity (as considered in Ref.~\cite{criger2018multi}), however belief-matching and belief-find can also handle more complicated error models arising from measurements in the syndrome extraction circuit.}
    \label{fig:belief_matching_illustration}
\end{figure*}

In this section, we first review the BP algorithm and then discuss how it is combined with minimum-weight perfect matching to exploit hyperedge error mechanisms when decoding circuit-level noise.

\subsection{Belief-matching and belief-find}

Our belief-matching and belief-find decoders are given a prior distribution of the error model (an assignment of an independent error probability to each of the edge or hyperedge error mechanisms), as well as the observed syndrome from the implemented error correction circuit.
Both decoders consist of two stages, illustrated in \Cref{fig:belief_matching_illustration} for the more simple case where syndrome measurements are perfect.

In the first stage, we use the BP algorithm to estimate a posterior distribution of the error model, given the observed syndrome.
More specifically, BP estimates the marginal probability that each possible error mechanism in the noisy syndrome extraction circuit has occurred (see \Cref{app:bp}).
Unlike a conventional MWPM or UF decoder, this stage uses knowledge of the full error model, including the hyperedge error mechanisms.
However, note that BP is only able to approximate the posterior distribution and does not have a threshold if used on its own, owing to the presence of short loops in the Tanner graph and degeneracy in the code~\cite{poulin2008iterative}.

In the second stage, we use the posterior marginal probabilities estimated by BP to set the edge weights in a matching graph.
This contrasts to a standard MWPM or UF decoder, where the prior distribution is used to set edge weights instead.
For surface codes, we can always decompose each hyperedge error mechanism $(t, u, v, w)$ into existing edges $(t, u)$ and $(v, w)$ in the matching graph, and the posterior marginal probability of each hyperedge is added to the marginal probabilities of the edges in its decomposition when setting edge weights.
After updating the edge weights, we decode the matching graph using MWPM~\cite{dennis2002topological} (for belief-matching) or weighted UF~\cite{delfosse2021almost, huang2020fault} (for belief-find).
See \Cref{app:belief_matching} for a more detailed description of belief-matching and belief-find.

We now consider the running time of belief-matching and belief-find.
The worst-case running time of belief-find is almost-linear in the number of error mechanisms, since the weighted UF decoder has almost-linear worst-case running time~\cite{delfosse2021almost,huang2020fault}, and BP has linear running time.
Furthermore, both weighted UF and the min-sum approximation of BP are comparatively simple decoding algorithms, which are amenable to implementation in hardware~\cite{das2020scalable,valls2021syndrome}.
For belief-matching, the worst-case running time is instead dominated by the MWPM step, which has worst-case running time $O(N^3\log(N))$, where $N$ is the number of nodes in the matching graph~\cite{higgott2021pymatching}.
However, the \textit{expected} running time of MWPM has been shown to scale approximately linearly with the number of error-detection events when below threshold~\cite{fowler2013minimum,higgott2023sparse,wu2023fusion}, and we have confirmed empirically that the expected running time of belief-matching is also approximately linear in this regime when using sparse blossom for the MWPM subroutine~\cite{higgott2023sparse}.
Furthermore, our numerical results demonstrate that the decoding performance of belief-find is almost identical to that of belief-matching, despite having significantly reduced worst-case running time. 
The BP step, although linear time, can still be quite computationally intensive, since the number of edges in the circuit-level Tanner graph is a constant factor larger than the number of edges in the corresponding matching graph, and running time does not depend strongly on the weight of the syndrome (it is not necessarily faster at low $p$, unlike MWPM or weighted UF).
However, we expect these challenges to be overcome since BP is highly parallelisable, and very fast implementations are already widely used for decoding classical LDPC codes.

Since the advantage that belief-matching and belief-find offer over MWPM or weighted UF alone derives from their use of \textit{hyperedges} present in the circuit-level Tanner graph, we expect them to outperform MWPM for most experimentally-relevant circuit-level noise models, for which the characterisation of hyperedge failure mechanisms is crucial to obtain good decoding performance~\cite{chen2021calibrated}.

\subsection{ML decoding with tensor networks}

We benchmark the performance of belief-matching against a maximum likelihood (ML) decoder for circuit-level noise, which outputs a Pauli correction that maximises the probability that the combined error and correction is in the stabilizer group.  Our ML decoder will use tensor network methods. We give the ML decoder the problem of decoding $L-1$ rounds of noisy stabilizer measurements, followed by a round of perfect stabilizer measurements.
After obtaining a set of noisy syndromes from the first $L-1$ rounds, an $n$-qubit Pauli error $E$ has accumulated on the code block from the execution of the measurement circuits. 
The final perfect round of syndrome measurement extracts the true syndrome of the error $E$.
Let $T$ be an $n$-qubit Pauli operator consistent with the true syndrome.
A circuit-level ML decoder finds a logical operator $\overline{L}\in\mathcal{C}(\mathcal{S})\setminus \mathcal{S}$ that maximises $\Pr([T\overline{L}])$, returning $T\overline{L}$ as the correction.
Here, the centralizer $\mathcal{C}(\mathcal{S})$ is the set of $n$-qubit Pauli operators that commute with all elements of $\mathcal{S}$ and the probability $\Pr([P])$ of the equivalence class $[P]\coloneqq \{PS: S\in\mathcal{S}\}$ is defined as $\Pr([P])\coloneqq \sum_{S\in\mathcal{S}}\Pr(PS)$, where $\Pr(P)$ is the probability that the error $P$ has accumulated on the code block, given the full syndrome and knowledge of the circuit-level noise model.

The ML decoder we implemented is the tensor network decoder described in Ref.~\cite{thomthesis}, which can be seen as a generalization of the BSV decoder of \citeauthor{bravyi2014efficient}~\cite{bravyi2014efficient} to the setting of imperfect syndrome measurements and circuit-level noise. 
The decoder is constructed by modelling all of the individual fault locations of the syndrome extraction circuit with individual tensors whose entries are probabilities of different Pauli errors having occurred, as defined in the circuit-level error model. By using the mathematical structure of a subsystem code called the \textit{circuit history code}~\cite{bacon2017sparse}, which is determined by
our syndrome extraction circuit, these individual fault tensors can be interconnected to a set of Kronecker delta tensors resulting in a tensor network which upon contraction allows us to find out the solution to the maximum-likelihood decoding problem.
Exact tensor network contraction, as with any approach to exact ML decoding, is computationally expensive and therefore slow. But the tensor network approach to ML decoding offers the advantage of being able to use numerical methods for 
approximate
tensor network contraction to lower the complexity of the calculation while maintaining a high degree of accuracy, which can be controlled.

\subsection{Performance of belief-matching and belief-find for depolarizing noise}

We compared the performance of belief-matching and belief-find to MWPM and union-find decoders through numerical simulations for the CSS surface code, using a standard circuit-level depolarizing noise model.
The details of the noise model are defined in \Cref{app:noise_model} (for which we set $\eta=1$ for depolarizing noise).
We used Stim to construct the detector error models, decompose hyperedges into edges and sample from the syndrome extraction circuits~\cite{gidney2021stim}. We used PyMatching to decode with MWPM~\cite{higgott2021pymatching}.
Throughout this work we estimate thresholds using the critical exponent method of Ref.~\cite{wang2003confinement}, with $1\sigma$ uncertainties in the last digit (estimated using jackknife sampling over lattice sizes) given in parentheses.

\begin{figure}
\centering
\includegraphics[width=0.8\columnwidth]{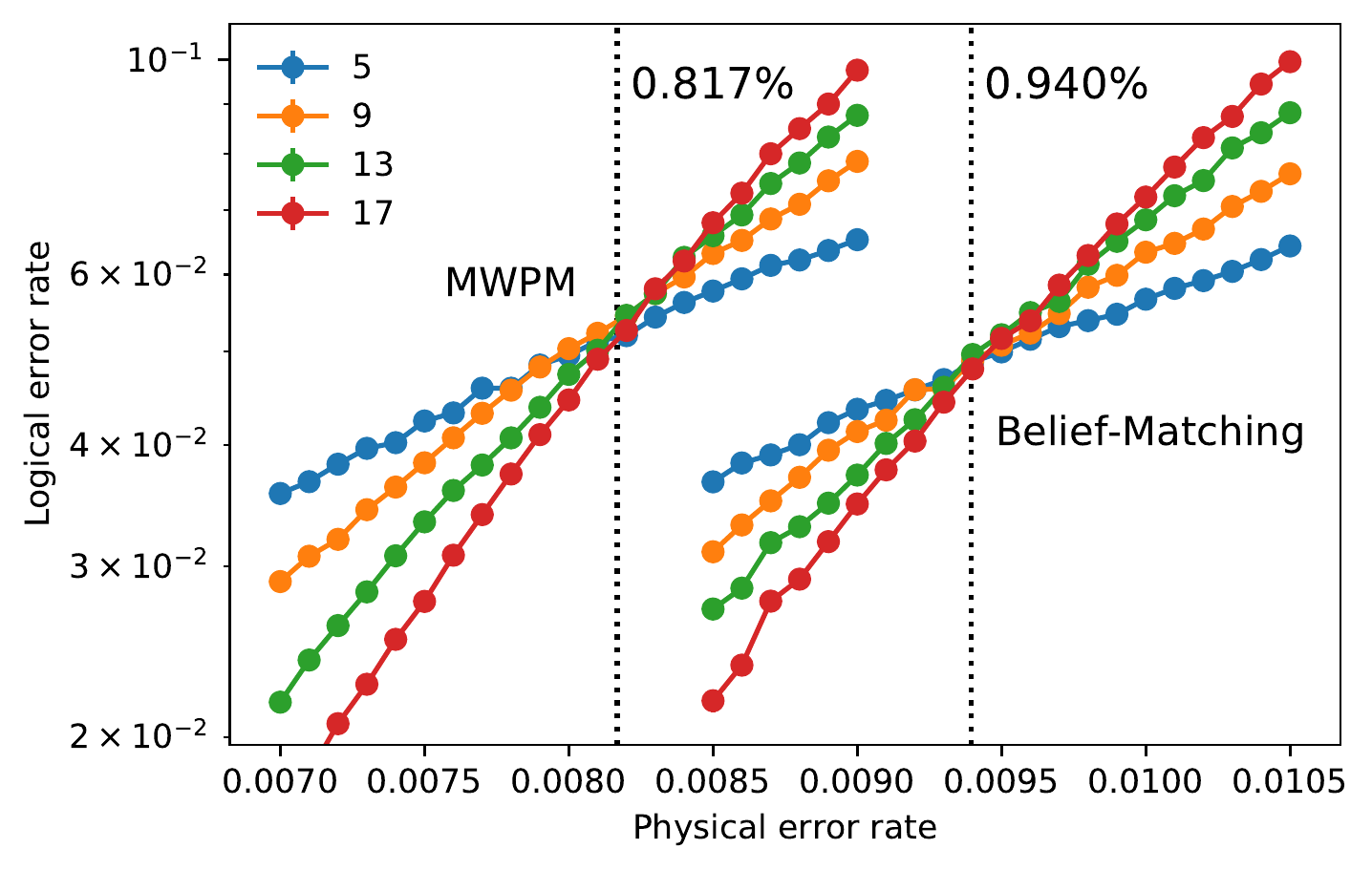}
\caption{Threshold of the square, CSS surface code using MWPM (left) and belief-matching (right) with $\eta=1$ depolarising noise. The logical failure rate is the probability that a logical $\overline{X}$ measurement outcome is flipped after preparing in an $\overline{X}$ eigenstate followed by $L$ rounds of stabilizer measurements.}
\label{fig:belief_matching_vs_mwpm_xz_code}
\end{figure}

\begin{figure}
\centering
\includegraphics[width=0.8\columnwidth]{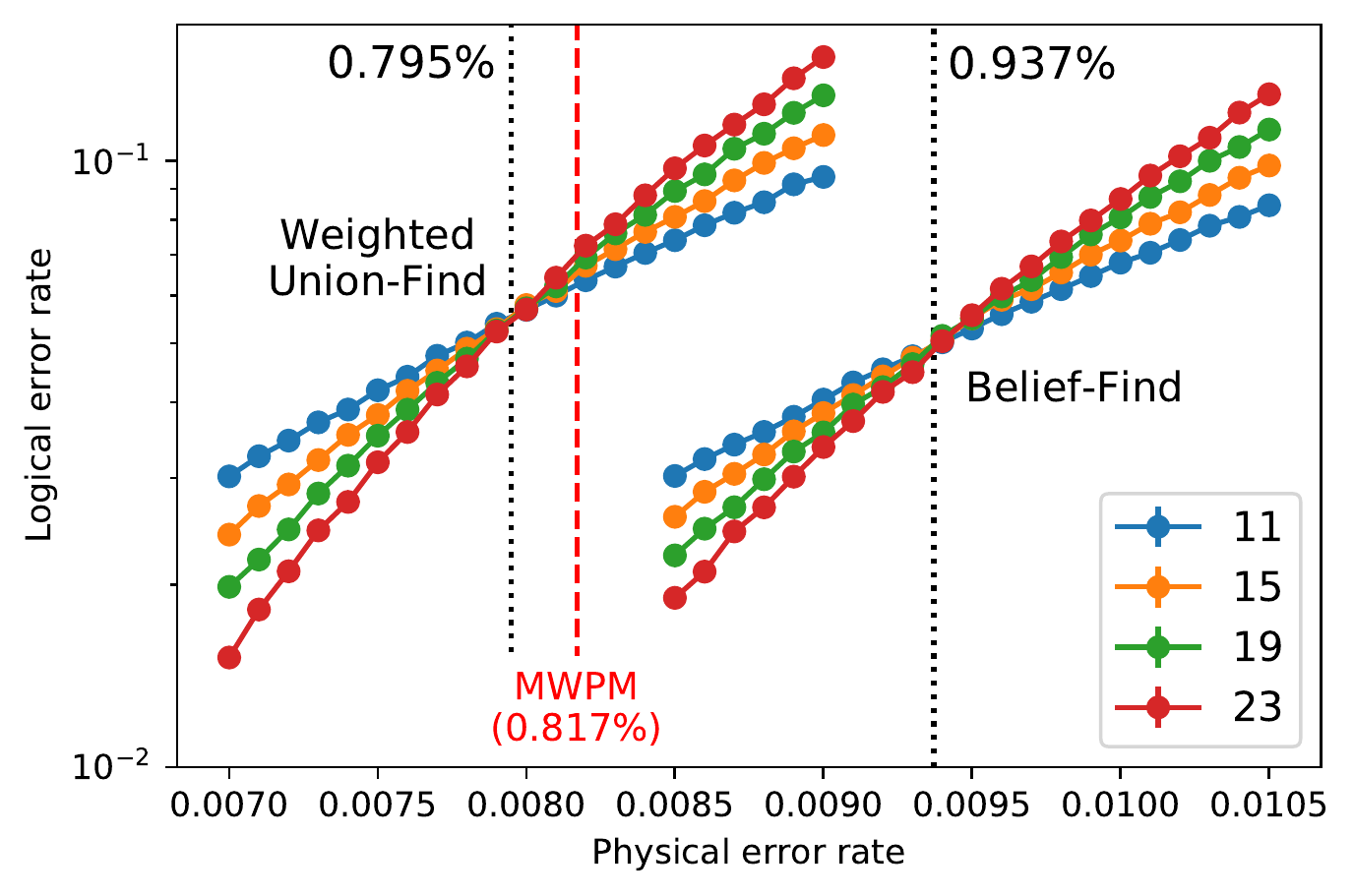}
\caption{Threshold of the square CSS surface code using weighted UF (left) and belief-find (right) for depolarising noise ($\eta=1$). The threshold value using MWPM for the same noise model is shown by the vertical red dashed line for comparison. The lattice size $L$ for each line is given in the legend, and the $x$ axis gives the physical error rate $p$ (equal to the CNOT infidelity for $\eta=1$). We find a threshold of 0.795(1)\% for weighted UF and 0.937(2)\% for belief-find.}
\label{fig:weighted_uf_vs_belief_find_threshold}
\end{figure}

In \Cref{fig:belief_matching_vs_mwpm_xz_code} we show the performance of belief-matching for the square, CSS surface code for circuit-level depolarising noise ($\eta=1$), and compare its performance to that of an uncorrelated MWPM decoder.
The MWPM decoder has previously had the highest reported circuit-level threshold for the surface code, which we find to be 0.817(5)\% for our noise model.
We find that belief-matching increases the threshold to 0.940(3)\%, a $1.15\times$ improvement.
This $1.15\times$ improvement can be attributed to belief-find taking advantage of correlations between the X and Z matching graphs due to Y errors.
\Cref{fig:weighted_uf_vs_belief_find_threshold} shows thresholds for circuit-level depolarising noise using the weighted UF decoder, as well as belief-find.
We find that belief-find also outperforms MWPM, achieving a threshold of 0.937(2)\% despite having a running time almost-linear in $N$.
We observe very little difference in decoding performance between weighted UF and MWPM alone, with weighted UF obtaining a threshold of 0.795(1)\%, compared to a threshold of 0.817\% for MWPM.
Furthermore, there is no statistically significant difference between the 0.940(3)\% threshold of belief-matching (see \Cref{fig:belief_matching_vs_mwpm_xz_code}) and the 0.937(2)\% threshold of belief-find.

\section{Tailoring codes to biased noise}
\label{sec:tailoring_codes}

Some physical systems can exhibit noise that is highly biased towards $Z$ errors~\cite{aliferis2008fault, puri2020bias, guillaud2019repetition, chamberland2020building}, for which the square CSS surface code is far from optimal.
Several proposals have been made for tailoring the surface code to these biased noise models.
One approach is to modify the basis of the stabilizer measurements in the surface code, while retaining the same square lattice layout. 
The XY surface code and XZZX surface code both follow this approach, and have been shown to have extremely high thresholds under biased noise~\cite{tuckett2018ultrahigh, tuckett2019tailoring, tuckett2020fault, bonilla2020xzzx, darmawan2021practical}.
Another approach is to apply schedule-induced gauge fixing to subsystem codes, which was shown to achieve high thresholds for the subsystem surface code in \cite{higgott2021subsystem} for biased noise.
Perhaps the simplest method of all is to modify the lattice dimensions of the CSS surface code, such that the $X$ distance and $Z$ distance are optimized for the bias~\cite{chamberland2020building,bonilla2020xzzx}.

Using our decoders, we tackle the problem of choosing an appropriate code in the presence of biased noise in planar architectures.
Inspired by the biased noise present in some quantum devices~\cite{chamberland2020building}, we consider a circuit-level noise model containing two parameters: a noise strength $p$ and a bias $\eta$.
The bias $\eta$ is the quotient of the probability that some $Z$-type error occurs, and the probability that any other occurs (for $P\in\{X,Y,Z\}$, a $P$-type Pauli operator on $n$ qubits is an operator in the set $\{I, P\}^{\otimes n}$).  See \Cref{app:noise_model} for more noise model details.
We are primarily interested in an optimization with respect to the required qubit overhead below threshold, in parameter regimes where useful fault-tolerant quantum computation is feasible. 
However, we do also compare the thresholds of the codes considered.

The variants of the surface code we study are the standard Calderbank-Shor-Steane (CSS) surface code, as well as the XY surface code, which uses $Y$-type stabilizers in place of $Z$-type stabilizers~\cite{tuckett2018ultrahigh, tuckett2019tailoring, tuckett2020fault}, both shown in \Cref{fig:Codes}.
For the CSS surface code, we allow the aspect ratio of the lattice to be optimized to reduce the qubit overhead below threshold.
For example, the $X$ distance can be reduced relative to the $Z$ distance for $Z$-biased noise (recall that the $P$ distance of a code is the minimum weight of a non-trivial $P$-type logical operator).
When the aspect ratio of a surface code is optimized in this way, we will refer to it as a \textit{rectangular} surface code.
For the XY surface code we consider only a square lattice geometry, since the aspect ratio here determines the $X$ and $Y$ distance, which should be equal for our chosen noise model in which $X$ and $Y$ errors are equiprobable.
The XY surface code has so far only been studied in an idealised setting of perfect syndrome measurements or a phenomenological noise model~\cite{tuckett2019tailoring,tuckett2020fault}.
To assess the practicality of the XY surface code for fault-tolerant quantum computing, it is important to study its performance for biased \textit{circuit-level} noise.
Furthermore, it is crucial to study how logical operations, such as logical state preparation, measurement, lattice surgery and magic state distillation can be implemented with the XY surface code, while still exploiting noise bias.

One reason the XY surface code is so promising for biased noise is that the $Z$ distance of the code is equal to the number of data qubits $n$, improving on the $O(\sqrt{n})$ $Z$ distance scaling of the square CSS surface code.
Furthermore, it was shown that under pure $Z$ noise, the code is equivalent to the repetition code, and therefore has a threshold of 50\%~\cite{tuckett2019tailoring}.
However, as we now show, these advantages of the XY surface code at infinite bias are \textit{fragile}, and can vanish at finite bias or at spatial or temporal boundaries.

\subsection{Fragility of the XY surface code}\label{sec:fragility}

\begin{figure}
    \centering
    
    \includegraphics[width=0.9\columnwidth]{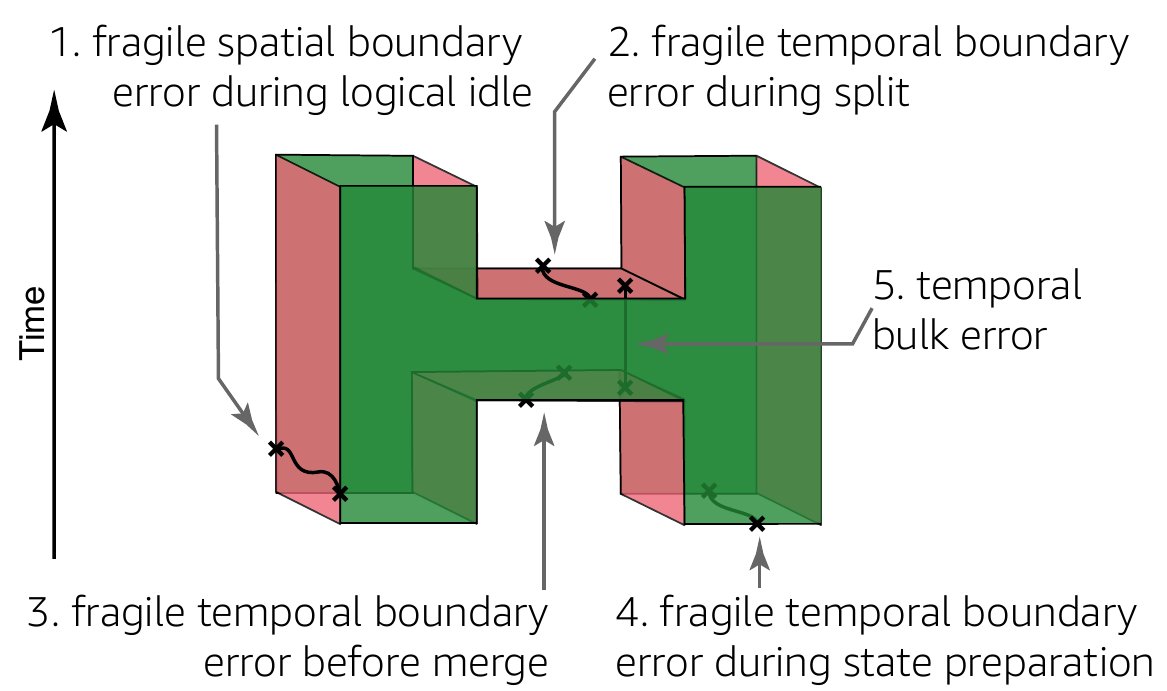}
    \caption{A space time diagram for two XY surface codes patches being prepared and then undergoing lattice surgery merging and splitting. We illustrate 5 different string-like errors that can lead to logical faults.
    Errors 1--4 are all purely boundary effects, constrained to either spatial (vertical) or temporal (horizontal) boundaries, and contain $O(\sqrt{n})$ $Z$ errors and at most one $X$ or $Y$ error. Error 5 is a sequence of measurement errors and we observe that these can also form strings between pairs of time-boundaries using $\tau$ measurement failures, where $\tau$ is the number of stabilizer rounds used during lattice surgery.}
    \label{fig:SpaceTimeDiagram}
\end{figure}

In this section, we will show that the protection provided by the XY surface code is \textit{fragile}, meaning that there are failure mechanisms in the XY surface code that require only $O(\sqrt{n})$ $Z$ errors during logical state preparation and measurement, as well as during logical idling at finite bias. 
An overview of all the errors discussed in this section is presented in \Cref{fig:SpaceTimeDiagram}.

\begin{figure}
\subfloat[\label{fig:finite_bias_z_error}]{\includegraphics[width=0.8\columnwidth]{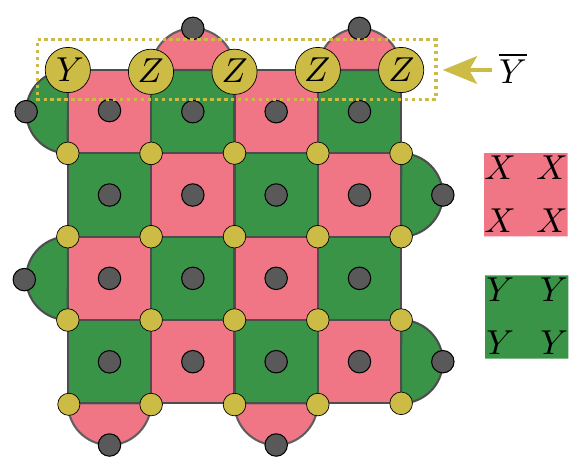}} \\
\subfloat[\label{fig:log_x_meas_error}]{
\includegraphics[width=0.8\columnwidth]{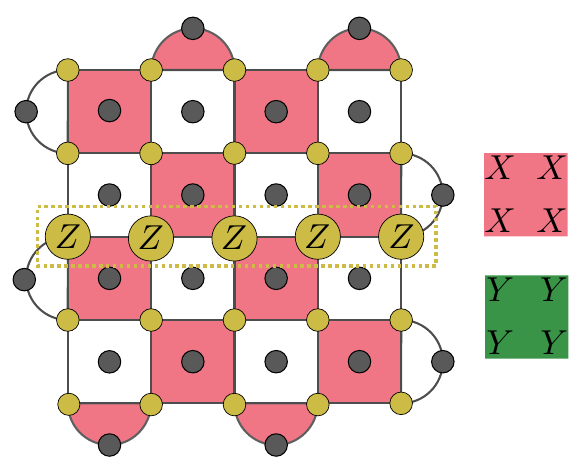}}
\caption{
Two types of fragile boundary errors.
(a) A fragile spatial boundary error that can occur at finite bias, involving a single $Y$ error, and $O(\sqrt{n})$ $Z$ errors.
(b) A fragile temporal boundary error.
It is a $Z$-type logical error with weight $O(\sqrt{n})$ that can occur during a logical $X$ measurement, when only $X$-type stabilizers are being measured.
}
\end{figure}

At finite bias, errors that consist of a mix of $X$, $Y$ and $Z$ Pauli operators are common.
We give examples of logical operators consisting of a single
$X$ or $Y$ error and $O(\sqrt{n})$ $Z$ errors.
An example of a logical $Y$ error of this form is shown in \Cref{fig:finite_bias_z_error}.
A similar logical $Y$ error can occur on the south boundary, and likewise logical $X$ errors consisting of a single $X$ and $O(\sqrt{n})$ $Z$ operators can occur on the east and west boundaries.
We will refer to any of these error patterns as \textit{fragile spatial boundary} errors as they only occur at planar code spatial boundaries and highlight the fragility of the infinite bias limit. In \Cref{fig:SpaceTimeDiagram}, error 1 is such an error. At low physical error rates and high bias, we would expect fragile spatial boundary errors to be dominant failure mechanisms.
Furthermore, since these failure mechanisms occur on all four boundaries, we expect a square aspect ratio to be optimal (assuming $X$ and $Y$ error probabilities are similar).
The existence of fragile spatial boundary errors emerges from the open boundary conditions.

Another example of fragility occurs during logical state preparation or measurement.
In order to measure the logical $X$ operator fault-tolerantly in the XY surface code, we measure all data qubits in the $X$ basis and infer both the $X$ logical operator and $X$ stabilizers in post-processing. Since the $X$ stabilizers are inferred from classical post-processing of data qubit measurements, rather than using an ancilla and measurement circuit, the $X$ stabilizers can be measured perfectly in this final round (and data qubit measurement errors can be interpreted as data qubit memory errors).
However, we cannot infer anything about the $Y$ stabilizers in this final round, since we measured the data qubits in the $X$ basis.
Since we measure only half of all the stabilizers, we no longer retain an $O(n)$ $Z$ distance at infinite bias. In \Cref{fig:SpaceTimeDiagram}, error 4 is such an error. In \Cref{fig:log_x_meas_error}, we show an example of an undetectable $O(\sqrt{n})$ $Z$-type logical failure mechanism that can occur just before (or during) a logical $X$ measurement, and which flips the outcome of the logical $X$ measurement.
The same type of fault can also occur during logical state preparation (e.g.~when preparing a logical $X$ eigenstate, data qubits are initialised in $\ket{+}$ states, and so only $X$ stabilizers can be measured initially).   Similarly, logical measurement and preparation in the $Y$ bias faces the same fragility problem.

Fragility also impacts lattice surgery, which is a computational primitive enabling surface code computation in a 2D layout~\cite{horsman2012surface,litinski2018lattice,litinski2019game,chamberland2021universal} through fault-tolerant measurements of logical multi-qubit Pauli operators.
In particular, fragile boundary errors can occur during the merge and split operations in lattice surgery and are equivalent to a logical idling error on some of the logical qubits.
Additionally, a string of measurement errors, which we refer to as temporal bulk errors, can lead to an incorrect measurement of the logical multi-qubit Pauli operator.
These string-like failure mechanisms are shown in \Cref{fig:SpaceTimeDiagram} and we descibe them in more detail in \Cref{app:latticeSurgery}.

\subsection{Below-threshold scaling of the XY surface code}\label{sec:performance_below_threshold}

We expect fragile boundary errors to have a significant impact on the performance of the XY code below threshold.
For simplicity, consider a noise model where the probability $p$ of a single-qubit $Z$ error is low, nevertheless it is substantially higher than the probability $p/\eta$ of a single-qubit $X$ or $Y$ error.
Temporal boundary errors are equivalent to the dominant failure mechanisms in the square CSS surface code, and decay as $O(p^{\sqrt{n}/2})$.
We expect fragile spatial boundary errors to decay as $O\left(p^{\sqrt{n}/2+O(1)}/\sqrt{\eta}\right)$ with minimum-weight decoding far below threshold.
To understand why this is the case, consider the most likely logical operator $E_d$ spanning the lattice that comprises $O(\sqrt{n})$ $Z$ errors and one $X$ or $Y$ errors.
We can split $E_d$ into two operators $E_a$ and $E_b$, where $E_a$ comprises one $X$ or $Y$ error and $c$ $Z$ errors and $E_b$ comprises $\sqrt{n}-1-c$ Z errors, where $c\in[0\twodots \sqrt{n}-1]$. We choose $c$ such that $E_a$ and $E_b$ both occur with probability $O\left(p^{\sqrt{n}/2+O(1)}/\sqrt{\eta}\right)$. Since $E_a$ and $E_b$ cannot be simultaneously correctable, the logical failure will be due to one of them occurring.
In most regimes of practical interest, we expect these string-like failure mechanisms to dominate over weight $n$ $Z$-type logical errors, which decay as $O(p^{n/2})$.
In order for weight $n$ $Z$-type logical errors to dominate we would expect $O(p^{n/2})\gg O\left(p^{\sqrt{n}/2+O(1)}/\sqrt{\eta}\right)$, which requires a bias $\eta\gg O((1/p)^{(n-\sqrt{n})})$.
However, the bias $\eta$ is a constant for any architecture, and so there will always be a value of $n$ above which string-like errors dominate.
As well as considering most-likely errors, it is important also to consider entropic contributions to the logical error rate, which are taken into account by our numerical simulations.
In our numerical simulations we analysed the decay in logical failure rate below threshold for a bias of $\eta=100$ (see \Cref{app:xy_below_threshold}).
As expected from the arguments in this section, we observe a decay of the form
$O(p^{\sqrt{n}/2}/\sqrt{\eta})$, instead of the $O(p^{n/2})$ scaling we might hope for at infinite bias (without SPAM errors).
In \Cref{app:state_prep_measure_numerics} we also provide numerical evidence that fragile temporal boundary errors lead to a significantly higher rate of errors during logical state preparation and measurement in the XY surface code.

\subsection{Decoder performance for biased noise}

\begin{figure}
\centering
\includegraphics[width=0.9\columnwidth]{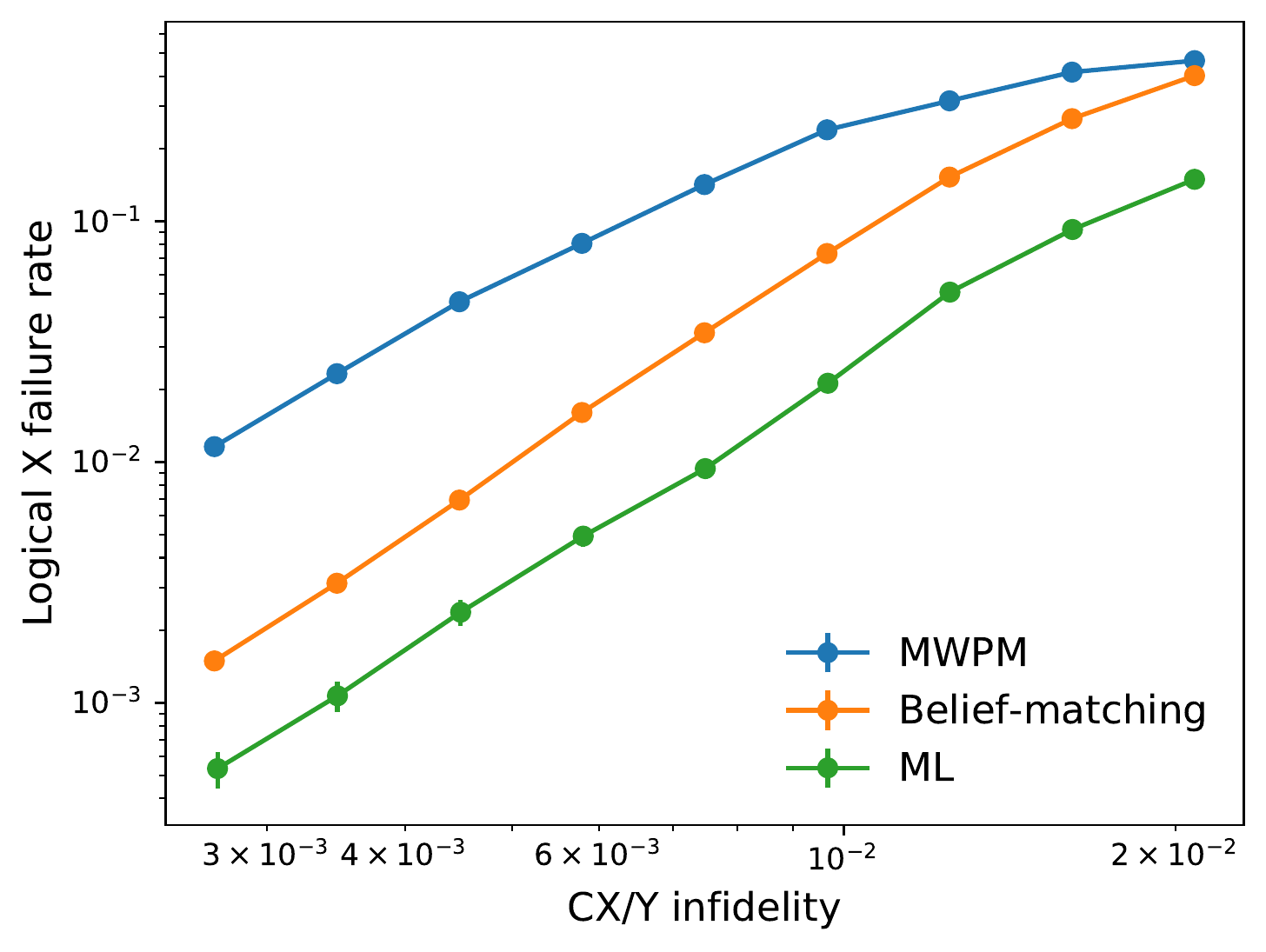}
\caption{Performance of the belief-matching decoder compared to a circuit-level ML decoder for
a $L=5$ XY surface code for $L$ rounds with perfect initialisation and syndrome measurements. We use the biased circuit level noise model defined in \Cref{app:noise_model} with $\eta=100$. Here we characterise the noise strength using the CNOT infidelity $p_{\mathrm{CX}}=(\frac{1}{5}+\frac{4}{5\eta})p$.
}
\label{fig:mwpm_vs_bp_mwpm_vs_ml}
\end{figure}

In \Cref{fig:mwpm_vs_bp_mwpm_vs_ml} we compare the logical error rate using the belief-matching decoder with that of pure MWPM and our circuit-level ML decoder, for an $L=5$ XY surface code for $L$ rounds with perfect initialisation and noisy syndrome measurements, with a bias of $\eta=100$.
At lower physical error rates (e.g.~$p=0.27\%$), we find that the logical error rate using belief-matching is around $7.8\times$ lower than MWPM alone, and 2.8$\times$ higher than ML decoding. 
The ML decoder was implemented in Julia using PastaQ~\cite{pastaq} to approximately contract the tensor network as a matrix product state, fixing a maximum bond dimension of $\chi=40$ throughout the contraction as we observed no further gains in accuracy by using a larger $\chi$.
These results show that belief-matching offers good performance relative to ML decoding for biased noise, despite having significantly reduced computational complexity.

\begin{figure}
\centering
\includegraphics[width=0.8\columnwidth]{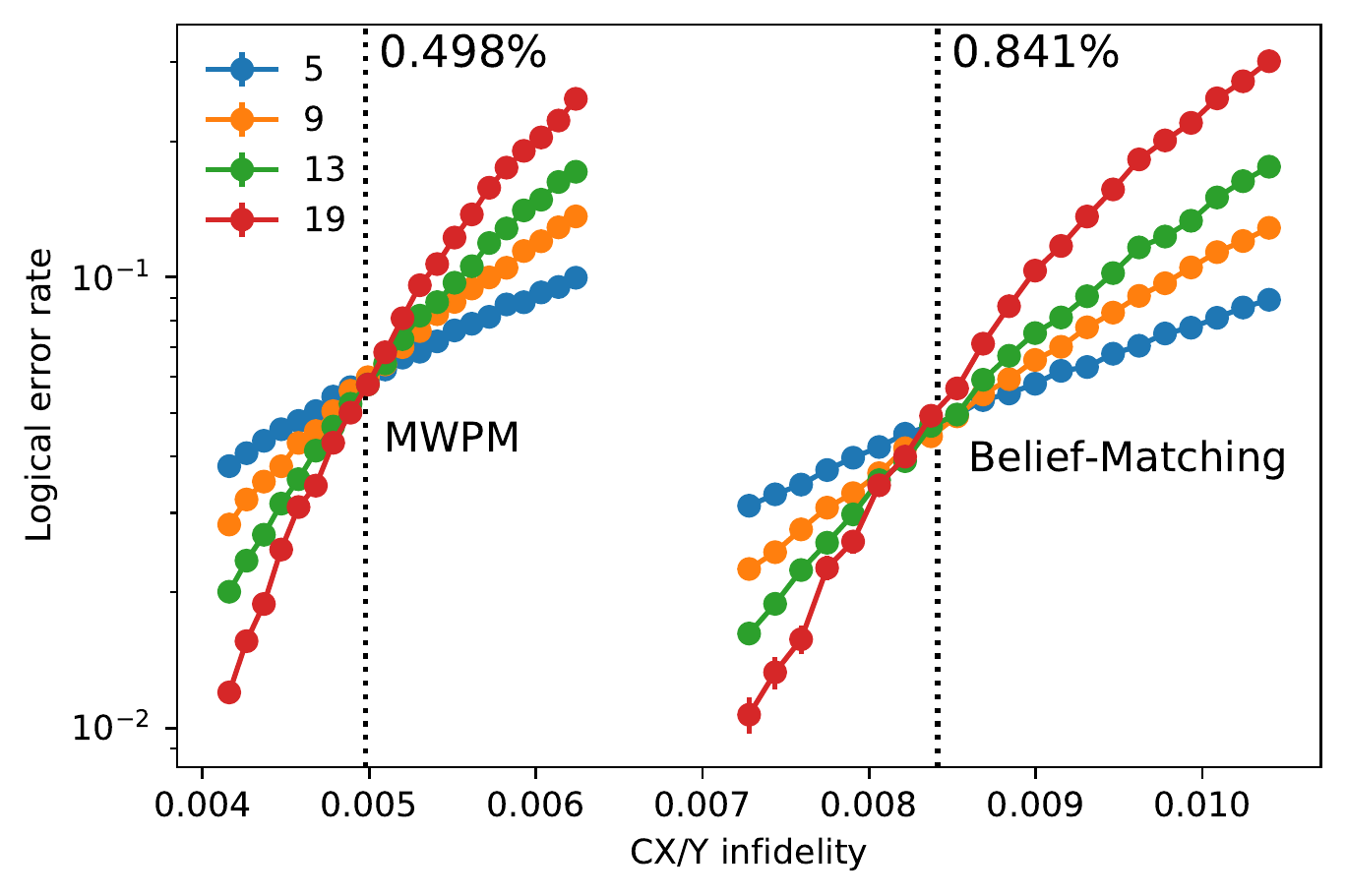}
\caption{Threshold of the XY surface code using MWPM (left) and belief-matching (right) for $\eta=100$. The lattice size $L$ for each line is given in the legend.}
\label{fig:tailored_biased_circuit_threshold}
\end{figure}

In \Cref{fig:tailored_biased_circuit_threshold} we compare the threshold of belief-matching with that of MWPM for the XY surface code for $\eta=100$ biased circuit-level noise.
We observe a threshold using belief-matching at 0.841(6)\% CNOT infidelity compared to 0.498(2)\% for MWPM, a $1.69\times$ relative improvement.

\subsection{Resource requirements of tailored surface codes}

\begin{figure}
\centering
\includegraphics[width=0.9\columnwidth]{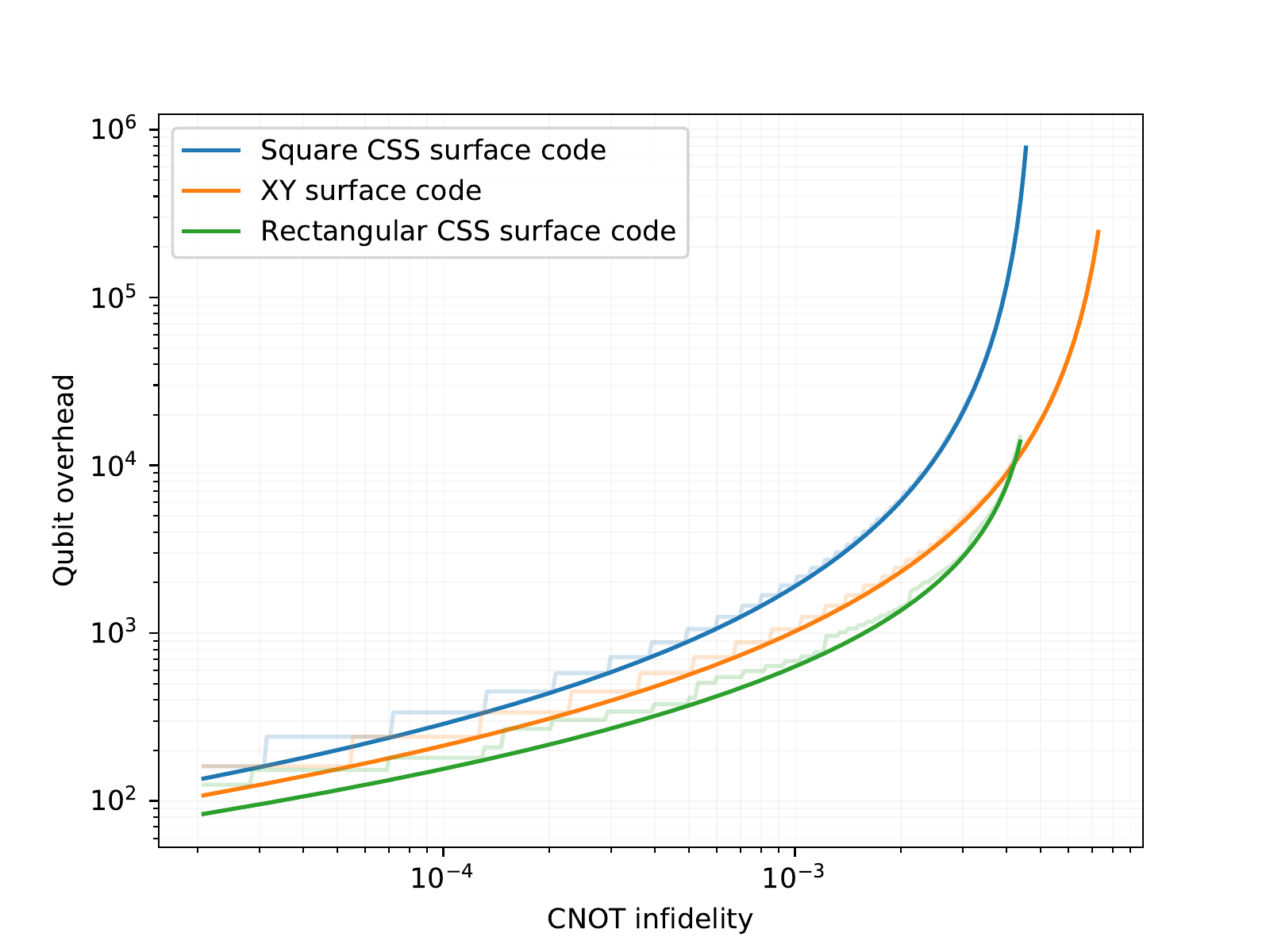}
\caption{Qubit overhead of the XY and CSS (square and rectangular) surface codes, as a function of CNOT infidelity, for the biased $\eta=100$ circuit-level noise model.
Translucent, stepped lines permit only odd, integer lattice sizes $L$, whereas smooth, solid lines interpolate and allow $L$ to be any real positive number. These overhead estimates were computed by solving the fitted ans\"atze in \Cref{eq:xy_ansatz}, \Cref{eq:xz_ansatz_1} and \Cref{eq:xz_ansatz_2} for the lattice dimensions, for a target logical error rate of $10^{-12}$.}
\end{figure}

Using our results for below-threshold scaling of the XY code, and by carrying out a similar analysis for the square and rectangular CSS surface codes, we compare the qubit overhead of the XY surface code with that of the square and rectangular CSS surface codes, to achieve a target logical error rate of $10^{-12}$ (the ``teraquop regime''~\cite{gidney2021fault}).
See \Cref{app:methods_for_numerics_tailored} for more details on how we carried out this analysis.

We find that the XY surface code outperforms the square CSS surface code in all regimes. However at physical error rates well below threshold, we find that the optimized aspect ratio of the rectangular CSS surface code allows it to outperform the XY surface code.
For instance, at a CNOT infidelity of $10^{-3}$, the XY surface code requires 1057 physical qubits per logical qubit, a substantial improvement on the 1921 required by the square CSS surface code.
However, by using the rectangular CSS surface code (optimizing the aspect ratio such that the $X$ and $Z$ logical error rates are approximately equal), we can achieve even better resource savings, requiring only 681 physical qubits per logical qubit to achieve the same logical error rate.

Therefore, although the XY surface code does still have a higher threshold, and improves over a CSS surface code of the same square aspect ratio, our results suggest that the ability to optimize the aspect ratio can be crucial for taking full advantage of the bias below threshold. 
Unfortunately, we do not expect to be able to improve the performance of the XY code by changing the aspect ratio, since the failure mechanisms described in \Cref{sec:fragility} can occur both horizontally and vertically.
However, we note that further improvements to the syndrome extraction circuits and decoding may lead to improved qubit overheads for the XY surface code.
Despite its good performance, we have shown that belief-matching does not match the performance of the computationally expensive maximum-likelihood decoder.
Therefore, future work could consider alternative decoders, with the hope of finding an efficient decoder with improved decoding performance that might lead to a more favourable qubit overhead for the XY surface code below threshold.

\section{Conclusion}
\label{sec:conclusion}

In this work we introduced new efficient decoders for the surface code, belief-matching and belief-find, which we showed have improved accuracy for decoding circuit-level noise.
Our decoders use knowledge of the full circuit-level noise model, i.e.~they consider all possible error mechanisms in the circuit along with their associated error probabilities.
By contrast, standard MWPM throws away much of the information contained in the circuit-level Tanner graph, since it only considers error mechanisms that are ``graphlike'' (errors that flip one or two detectors).
We therefore expect that belief-matching and belief-find will have good performance for a wide range experimentally-relevant noise models and can use noise models calibrated from experimental data~\cite{chen2021calibrated}.
Indeed, after our pre-print was released, our belief-matching decoder was used to experimentally demonstrate the suppression of quantum errors by scaling a surface code logical qubit from distance 3 to 5~\cite{google2023suppressing}.
In this surface code experiment, it was shown that belief-matching outperformed both MWPM~\cite{dennis2002topological,fowler2013minimum} and the correlated MWPM decoder of~\cite{fowler2013optimal} for experimental noise~\cite{google2023suppressing}.
The improved accuracy relative to the correlated MWPM decoder of Ref.~\cite{fowler2013optimal} can be understood from the fact that belief-matching considers the full circuit-level noise model, whereas correlated MWPM considers each pair of correlated edges in isolation and only updates edge weights in close proximity to an initial (uncorrelated) MWPM solution.
Our belief-find decoder has an almost-linear worst-case runtime while having very similar accuracy to belief-matching.
This worst-case runtime is a significant improvement on the worst-case runtimes of MWPM and belief-matching, although these matching decoders can still have a linear expected runtime in practice at low error rates.
Future work could explore implementations of belief-matching and belief-find in hardware~\cite{das2020scalable, valls2021syndrome}.
Belief-matching and belief-find can be applied to any code for which MWPM can be used, which includes 2D surface codes~\cite{dennis2002topological,tuckett2018ultrahigh,bonilla2020xzzx,breuckmann2016constructions,dua2022clifford}, subsystem surface codes~\cite{bravyi2012subsystem,higgott2021subsystem} and Floquet codes~\cite{hastings2021dynamically}, amongst others~\cite{bacon2006operator,li20192d,chamberland2020topological}.
Previous work developing decoders that handle hyperedge error mechanisms in the surface code have mostly assumed perfect syndrome measurements or a phenomenological error model~\cite{duclos2010fast,duclos2010renormalization,duclos2013fault,wootton2012high,hutter2014efficient,bombin2012strong,bombin2012strong,criger2018multi}.
More generally, we have demonstrated how high performance decoders for classical LDPC codes (such as BP) can be applied directly to infer probable error locations in realistic circuit-level noise models, and expect that our work will inspire the application of similar techniques to other quantum error correction codes and protocols.

As an application of our decoders, we have also investigated the performance of the XY surface code for fault-tolerant quantum computation in the presence of biased noise.
Although the XY surface code has a $Z$ distance of $n$, we have identified string-like failure mechanisms, which we refer to as fragile boundary errors, that can occur at temporal boundaries (during logical state preparation and measurement), or at spatial boundaries at finite bias.
These fragile boundary errors consist of $O(\sqrt{n})$ $Z$ errors and $O(1)$ $X$ or $Y$ errors, and will likely dominate over errors due to the weight $n$ $Z$-type logical in most realistic settings.
We showed that belief-matching has good performance for handling biased circuit-level noise, and used it to benchmark the performance of the XY surface code compared to the CSS surface code, for which the lattice dimensions can be tailored to the bias.

There are other proposals for handling biased noise which we have not considered in this work.
The XZZX surface code is a promising candidate, which has been shown to achieve very high thresholds in the presence of biased noise~\cite{bonilla2020xzzx,darmawan2021practical}.
However, optimizing the dimensions of the XZZX surface code requires using an unrotated geometry, which requires $2\times$ more qubits than the CSS surface code to achieve the same distance.
Another option is to use the subsystem surface code (SSC)~\cite{bravyi2012subsystem} with schedule-induced gauge-fixing, which has also been shown to have high thresholds for biased circuit-level noise~\cite{higgott2021subsystem}.
The SSC uses $1.75\times$ more qubits than the CSS surface code to achieve the same distance (assuming one ancilla per gauge operator), but may be easier to build owing to its weight-three checks and reduced connectivity requirements, which could reduce crosstalk and frequency collisions~\cite{chamberland2020topological}.
Importantly, unlike the XY surface code, the aspect ratios of the CSS, XZZX and subsystem surface codes can all be optimized in the presence of bias, which we have shown is highly desirable for reducing qubit overheads.
While the XZZX and subsystem surface codes both offer improved performance compared to the CSS surface code near threshold for biased circuit-level noise~\cite{darmawan2021practical, higgott2021subsystem}, a more detailed analysis will be required to assess whether this also translates into a reduced qubit overhead for a noise regime of practical interest below threshold.
The CSS, XY and XZZX surface codes all fall within the broader family Clifford-deformed surface codes~\cite{dua2022clifford}, and we even proposed a new Clifford-deformed code in \cref{sec:mitigating_fragile_errors}. These provide even more flexibility for tailoring the surface code to the noise bias, and further work is required to investigate these codes in a fault-tolerant setting and our decoders provide a powerful tool to enable this further research.
Finally, for architectures with improved qubit connectivity, it is possible that bias-tailored quantum LDPC codes will offer a further reduction in qubit overhead~\cite{roffe2022bias}.

\section{Contributions}

OH developed the belief-matching and belief-find algorithms and ran most of the numerical simulations with supervision from ETC. TCB performed the ML numerics with supervision from STF and AK. OH and ETC identified the fragile boundary errors. OH, ETC, AK and STF contributed to the analysis in \Cref{sec:mitigating_fragile_errors}. All authors contributed to writing the manuscript.

\section{Acknowledgements}

This work was initiated when OH, TB and ETC worked at Amazon Web Services. OH acknowledges support from the Engineering
and Physical Sciences Research Council [grant number EP/L015242/1] and a Google PhD fellowship. OH would like to thank Nikolas Breuckmann, Christopher Chamberland and Michael Newman for insightful discussions. 
We thank Ben Brown, Neil Gillespie and Luigi Martiradonna for providing helpful feedback on the manuscript. 

\bibliography{references.bib}

\appendix

\section{BP review}\label{app:bp}
The BP algorithm, also known as the \textit{sum-product algorithm}, is an efficient iterative message passing algorithm with good performance for decoding classical low-density parity check (LDPC) codes~\cite{mackay1996near}.
Consider a binary check matrix $H$ defining a linear code $\mathrm{ker}(H)$.
BP is most readily understood by considering the \textit{Tanner graph} $\mathcal{T}(H)$ of the check matrix $H$.
The Tanner graph is a bipartite graph with a \textit{check node} for each parity check (row of $H$), and a \textit{variable node} for each bit (column of $H$), and graphically represents a factorisation of the joint probability distribution over the bits.
Each check node is connected by an edge to the variable nodes corresponding to the bits it acts nontrivially on.
The BP algorithm takes as input the prior probabilities that each bit is flipped, as well as the syndrome of each parity check.
Each iteration of BP consists of a \textit{horizontal} step and a \textit{vertical} step.
In the horizontal step, each check node (a row of $H$) sends a message to its adjacent variable nodes. In the vertical step, each variable node (a column of $H$) sends a message to its adjacent check nodes, where each message is essentially a local application of Bayes' rule.
In each iteration, the latest check-to-variable messages can be used, along with the priors, to compute the `pseudoposterior probabilities', which approximate the marginal probabilities that each bit has been flipped, given the priors and the syndrome.

To improve the numerical stability and efficiency of BP, we use log-likelihood ratios (LLR) to represent probabilities and compute messages, where the LLR of a binary random variable $U$ is defined as
\begin{equation}
L(U) = \log\left[\Pr(U=0)/\Pr(U=1)\right].
\end{equation}
We denote by $q_i$ the LLR of the pseudoposterior probability that bit $i$ was flipped and define a binary vector $\mathbf{x}$ of \textit{hard decisions} where element $\mathbf{x}[i]$ is set to $0$ if $q_i>0$ and is set to $1$ if $q_i\leq 0$.
In each iteration of BP we compute $H\mathbf{x}$, and stop the algorithm and return $\mathbf{x}$ if $H\mathbf{x}=\mathbf{s}$, where $\mathbf{s}$ is the syndrome.
When this happens we say that BP has \textit{converged}.
If a maximum number of iterations $m_\text{iter}$ is reached without BP converging, then we record a heralded failure (and we set $m_\text{iter}=30$ in this work).
We refer the reader to Refs.~\cite{mackay2003information,chen2005reduced} for a more detailed overview of BP and its variants.

While BP is an effective decoder for classical LDPC codes, its application to quantum codes faces challenges.
Most notably, the marginals output by BP cannot be used to distinguish between multiple equiprobable solutions to the decoding problem that differ by stabilizers~\cite{poulin2008iterative}.
Several modifications of BP have been used with the purpose of fixing the problem that quantum degeneracy poses for the BP decoder, most notably the use of \textit{ordered statistics decoding} (OSD) post-processing of the BP posterior marginal probabilities~\cite{fossorier2001iterative}, which was successfully used to decode hypergraph product codes in Refs.~\cite{panteleev2019degenerate,roffe2020decoding,roffe2022bias}.

\section{The circuit-level Tanner graph}\label{app:circuit_tanner}

We define a Tanner graph describing the circuit-level noise model, which we call the circuit-level Tanner graph $\mathcal{T}_{\mathrm{CL}}=(V,C,E)$, where $V$ is a set of \textit{variable nodes}, $C$ is a set of \textit{check nodes} and $E$ is the edge set.
Recall that a Tanner graph is a bipartite graph, so for each edge $(v, c)\in E$ we have $v\in V$ and $c\in C$.
Each check node $c\in C$ corresponds to a \textit{detector}~\cite{gidney2021stim} and each variable node $v\in V$ corresponds to an error mechanism that can occur in the syndrome extraction circuit.
There is an edge $(v, c)\in E$ if and only if the error mechanism corresponding to $v\in V$ flips the detector corresponding to $c\in C$.
If multiple error mechanisms trigger the same set of detectors (and are thus indistinguishable), then these error mechanisms are merged into a single variable node which is assigned a probability equal to the probability that an odd number of the errors occurred. A variable node is said to flip if the error mechanism it corresponds to occurs in the circuit (or, if it corresponds to multiple equivalent error mechanisms, then it flips if an odd number of these errors occur).
The set of prior probabilities $p_{\mathrm{prior}}$ includes, for each variable node $v$, the probability $p_{\mathrm{prior}}^v$ that it would flip under the noise model.
A circuit-level Tanner graph is a graphical representation of a \textit{detector error model} in Stim~\cite{gidney2021stim}, but where equivalent variable nodes have been merged as just described.

\Cref{fig:circuit_level_tanner_graph} shows a Tanner graph for a biased circuit-level noise model, restricted to the 15 non-trivial two-qubit Pauli errors that can occur after a single CNOT gate in the parity check measurement schedule (it is a small subgraph of the full circuit-level Tanner graph).
A Tanner graph describes a factorisation of a joint probability distribution in which each bit (corresponding to a variable node) is flipped \textit{independently} with the assigned prior probability.
Note that, in the standard Pauli circuit noise models considered in the literature and in this work, the probabilities of each Pauli error that can occur after a gate are described as probabilities of disjoint errors, rather than as independent events.
While some specific Pauli noise models, such as the depolarising noise model, \textit{can} be described as an independent distribution~\cite{chao2020optimization}, this is not the case in general.
When computing priors and constructing the Tanner graph, we make the approximation that each probability of a disjoint error mechanism instead corresponds to the probability of an independent error mechanism.
This approximation is correct to leading order in $p$, and therefore a good approximation for the physical error rates we considered.

\section{The belief-matching and belief-find decoders}
\label{app:belief_matching}

We can run BP directly on the circuit-level Tanner graph in order to estimate the marginal probability that each error mechanism occurred.
However, due to low weight degenerate errors and loops in the Tanner graph, BP on its own is known not to have a threshold for the surface code with perfect syndrome measurements, and we confirmed that this is also the case with circuit-level noise.
We instead use the BP posteriors to choose edge weights for a matching graph $\mathcal{G}$, which we can decode using a toric code decoder that handles weighted edges, such as MWPM or weighted UF~\cite{delfosse2021almost,huang2020fault}.
Each node in $\mathcal{G}$ either corresponds to a detector or is a boundary node.
There is an edge $(u, v)$ in $\mathcal{G}$ for each variable node of degree one or two in the circuit-level Tanner graph.
If a variable node $m$ in $\mathcal{T}_{\mathrm{CL}}$ has degree two, then $u$ and $v$ in the corresponding edge $(u,v)$ in $\mathcal{G}$ are the two detectors that $m$ is adjacent to in $\mathcal{T}_{\mathrm{CL}}$.
If a variable node $m$ in $\mathcal{T}_{\mathrm{CL}}$ instead has degree one, then the corresponding edge $(u, v)$ in $\mathcal{G}$ consists of the detector $u$ that $m$ is adjacent to in $\mathcal{T}_{\mathrm{CL}}$, as well as a boundary node $v$.
The circuit-level Tanner graph for the XY (and CSS) surface code also contains variable nodes with degree greater than two, which would correspond to hyperedges in a decoder hypergraph (the obvious generalisation of a matching graph). 
However, for the surface code, these hyperedges can always be approximated by a sets of edges already present in the matching graph~\cite{gidney2021stim}.
For example, consider a weight-four hyperedge $h\coloneqq(t,u,v,w)$, and assume that the edges $e_1\coloneqq(t, u)$ and $e_2\coloneqq(v, w)$ are already present in the matching graph, we say that the hyperedge $h$ can be decomposed into the edges $e_1$ and $e_2$.
Let $p_{\mathrm{BP}}(v)$ be the marginal posterior probability output by BP for variable node $v$ (an edge or hyperedge), and let $D(e)$ be the set of all hyperedges which have 
$e$ in their decomposition.
For each edge, we define an adjusted probability 
\begin{equation}\label{eq:edge_weight}
p_{\mathrm{adj}}(e)\coloneqq p_{\mathrm{BP}}(e) + \sum_{h\in D(e)}p_{\mathrm{BP}}(h)
\end{equation}
and set $p_w(e)\coloneqq\min(p_{\mathrm{adj}}(e), 1)$~\footnote{Alternatively we could treat $p_{\mathrm{BP}}$ as probabilities of independent events (even though they are not), and take $p_w(e)$ to be the probability that an odd number of faults in the set $\{e\}\cup D(e)$ occurred under this assumption. This alternative approach would be consistent with how edges and hyperedges are usually merged in matching graphs.}.
We then assign the weight $w(e)\coloneqq -\log(p_w(e))$ to each edge in the matching graph~\footnote{Another natural choice of edge weight would be to use $w(e)\coloneqq \log((1-p_w(e))/p_w(e))$ and then handle the negative weights with MWPM or weighted UF using the method in \cite{higgott2021pymatching}. However, we find that our choice $w(e)\coloneqq -\log(p_w(e))$ instead leads to slightly improved decoding performance for both belief-matching and belief-find}.
Note that we always ensure that each hyperedge has a unique decomposition into edges (if there exists more than one valid decomposition, then we pick one arbitrarily).

Once we have used BP to construct the matching graph $\mathcal{G}$, we use either MWPM or weighted UF to decode it.
When we use MWPM as a subroutine, we refer to our decoder as \textit{belief-matching}, and when we instead use weighted UF as a subroutine, we refer to it as \textit{belief-find}.
The MWPM decoder finds a set of edges in $\mathcal{G}$ consistent with the syndrome that have minimal total weight, and a standard exact implementation of the algorithm has a worst case running time of $O(N^3\log(N))$, where $N$ is the number of nodes in $G$~\cite{dennis2002topological,edmonds1965paths,higgott2021pymatching}.
In this work, we use the PyMatching implementation of MWPM~\cite{higgott2021pymatching}.
Weighted UF instead finds a low-weight (but in general not minimal weight) solution, but has an almost-linear worst case running time of $O(N\alpha(N))$, where $\alpha$ is the inverse of Ackermann's function, which grows very slowly~\cite{delfosse2021almost,huang2020fault}.
While the original UF algorithm did not use the weights of edges in $\mathcal{G}$~\cite{delfosse2021almost}, it was shown in Ref.~\cite{huang2020fault} that using the edge weights during the ``cluster growth'' stage (also called the \textit{syndrome validation} stage) led to significantly improved decoding performance, while maintaining the same asymptotic running time (for edge weights of some fixed precision).
Our implementation of weighted UF is very similar to the version used in Ref.~\cite{pattison2021improved}.
As in Ref.~\cite{pattison2021improved}, we grow clusters on a \textit{split-edge} graph $H$, obtained from $G$ by adding a node in the middle of each edge.
We find that this modification significantly improves decoding performance.
Additionally, in each round of growth, we grow smaller odd clusters before larger ones and fuse clusters at the endpoints of an edge (and update their parity) as soon as the edge becomes fully grown.
This means we do not grow a cluster if its parity has already changed from odd to even earlier in the same round of growth (unlike in Algorithm 2 of Ref.~\cite{delfosse2021almost}).
Finally, we construct a spanning tree, not a minimum-weight spanning tree, in the peeling decoder stage of weighted UF (here we are consistent with Ref.~\cite{delfosse2021almost} but not Ref.~\cite{huang2020fault}).
None of these modifications affect the asymptotic running time of the algorithm.
In our implementation, we only decode the matching graph using MWPM or weighted UF on instances where BP alone does not converge (almost all failures for BP alone are due to the algorithm not converging).
Our belief-matching and belief-find decoders are summarised in \Cref{algo:belief_matching}

\begin{algorithm}[H]
\caption{Belief-matching/ belief-find}
\label{algo:belief_matching}
\begin{algorithmic}[1]
\REQUIRE The circuit-level Tanner graph $\mathcal{T}_{\mathrm{CL}}$, the priors $p_{\mathrm{prior}}$ and the syndrome $\sigma$
\ENSURE A correction operator, given as a set of variable nodes in $\mathcal{T}_{\mathrm{CL}}$
\STATE Compute the marginal posterior probability $p_{\mathrm{BP}}(v)$ for each variable node $v$ by running BP, which takes $\mathcal{T}_{\mathrm{CL}}$, $p_{\mathrm{prior}}$ and $\sigma$ as input.
\STATE Find a tentative correction $c^\prime$, which is the set of variable nodes $v$ for which $p_{\mathrm{BP}}(v)>0.5$. We say that BP has \textit{converged} if $c^\prime$ also has syndrome $\sigma$.
\IF {BP has converged} 
    \RETURN The set of variable nodes $c^\prime$
\ELSE
	\STATE Distribute the posterior $p_{\mathrm{BP}}(h)$ of each hyperedge $h$ to the edges in its decomposition and, using \Cref{eq:edge_weight}, compute the edge weights in the matching graph $\mathcal{G}$.
	\STATE Decode $G$ with syndrome $\sigma$ using MWPM (for belief-matching) or weighted UF (for belief-find), to find a set of edges $E$
	\RETURN The variable nodes in $\mathcal{T}_{\mathrm{CL}}$ corresponding to the edges $E$
\ENDIF
\end{algorithmic}
\end{algorithm}

\section{Noise model used in numerical simulations}\label{app:noise_model}

For our numerical simulations, we used the same biased circuit-level noise model as in Ref.~\cite{chamberland2021universal} that is captured by two parameters $p$ and $\eta$; there are, however, alternative definitions~\footnote{The parameters $p$ and $\eta$ describing the biased circuit-level noise can be alternatively defined as follows: $p$ is the probability of any error at the given location, whereas $\eta$ is the ratio of the probabilities for $Z$ errors and any other errors.
For instance, for single-qubit locations we would have
$p = p_X + p_Y + p_Z$ and $\eta = p_Z /(p-p_Z )$, where $p_P$ is the probability of a single-qubit $P$ error.
}.
Namely, each two-qubit gate is followed by a two-qubit Pauli channel, for which $ZZ$, $ZI$ or $IZ$ can occur with probability $p/15$ each, and the remaining 12 non-trivial two-qubit Paulis can each occur with probability $\frac{p}{15\eta}$.
Each single qubit gate location or single qubit idle location of the same duration (a single time step) is followed by a $Z$ error with probability $p/3$ or an $X$ or $Y$ error each with probability $\frac{p}{3\eta}$.
A $\ket{+}$ state is incorrectly prepared as a $\ket{-}$ state with probability $2p/3$, and a $\ket{0}$ state is incorrectly prepared as a $\ket{1}$ state with probability $\frac{2p}{3\eta}$.
Each single-qubit $X$-basis measurement is flipped with probability $2p/3$, and each single-qubit $Z$-basis measurement is flipped with probability $\frac{2p}{3\eta}$.
We use the CNOT infidelity $p_{\mathrm{CX}}$ when determining and comparing thresholds since it is a useful measure of the noise strength; note that the parameter $p$ only corresponds to the CNOT infidelity for $\eta=1$, since $p_{\mathrm{CX}}=(\frac{1}{5}+\frac{4}{5\eta})p$.
Each single-qubit gate and two-qubit gate has a duration of a single time step, whereas single-qubit state preparation and measurement are each taken to have a duration of half a time step.
For the XY code measurement schedule we used, $X$ stabilizers are measured using CNOT gates controlled on an ancilla initialised in a $\ket{+}$ state, and $Y$ stabilizers are measured using controlled-Y (CY) gates, also controlled on a $\ket{+}$ state.
These two-qubit gates are applied in the order indicated by the blue text in \Cref{fig:circuit_level_tanner_graph}.
For the CSS surface code we used the same schedule for measuring $X$ stabilizers, and $Z$ stabilizers were measured using CNOT gates targeted on an ancilla initialised in the $\ket{0}$ state, and applied in the same order as used for CY gates in the XY surface code schedule.
We assume that CY gates can be implemented natively, with the same noise model as CX gates.
Other than for \Cref{fig:spam_vs_no_spam}, where we analyse the effect of fragile temporal boundary errors, we assume perfect state preparation and measurement for all other numerical simulations (a perfect round of stabilizer measurements is inserted after perfect initialisation of data qubits, and before perfect logical measurement of the data qubits).
In general, we have made optimistic assumptions for our XY surface code simulations (perfect logical initialisation and measurement, native CY gates), in order to understand if fragile spatial boundary errors alone result in inferior performance relative to a rectangular CSS surface code. Removing these optimistic assumptions will only make performance of the XY surface code worse.

\section{Fragility of lattice surgery}\label{app:latticeSurgery}

\begin{figure}
\includegraphics[width=0.9\columnwidth]{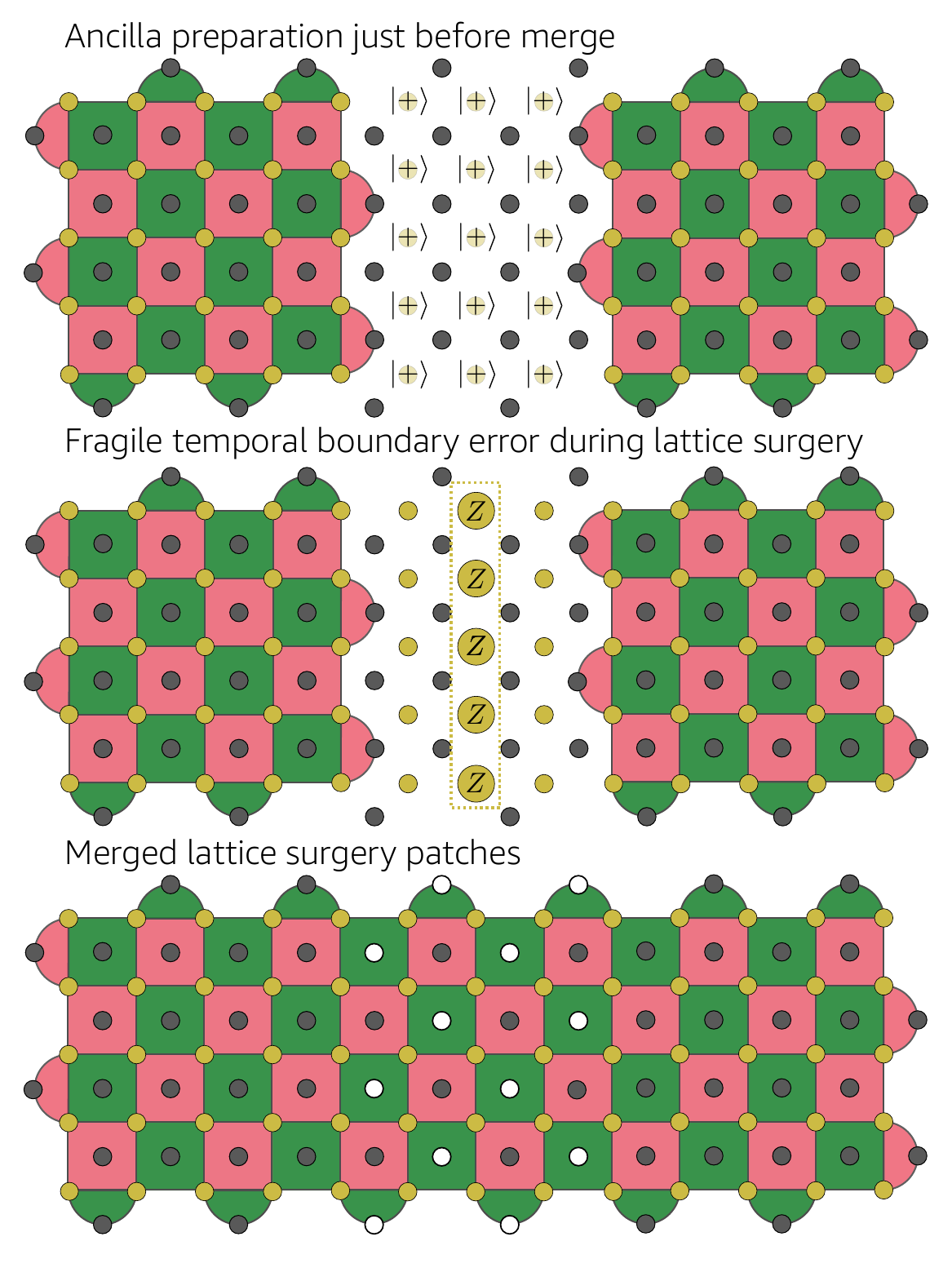}
\caption{We illustrate the merge step of lattice surgery performing a logical $\overline{Y} \otimes \overline{Y}$ measurements between two square XY surface code patches, including a possible fragile temporal boundary error.  Before the merge, the data qubits between patches must be prepared in the $\vert + \rangle$ state.  We illustrate a possible fragile temporal boundary error that occurs during or after the $\vert + \rangle$ state preparation but before the merge stabilizers are measured.  During the merge step, the vertices highlighted white have random outcomes except that their product gives the outcome of the logical $\overline{Y} \otimes \overline{Y}$ measurement.}
\label{fig:LatticeSurgery}
\end{figure}

\begin{figure}[ht!]
\centering
\subfloat[\label{subfig:tailored_layout_ETC}]{
\includegraphics[width=0.8\columnwidth]{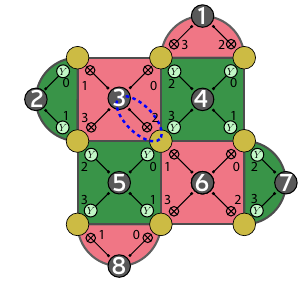}
}\hfill
\subfloat[\label{subfig:circuit_tanner}]{
\includegraphics[width=0.9\columnwidth]{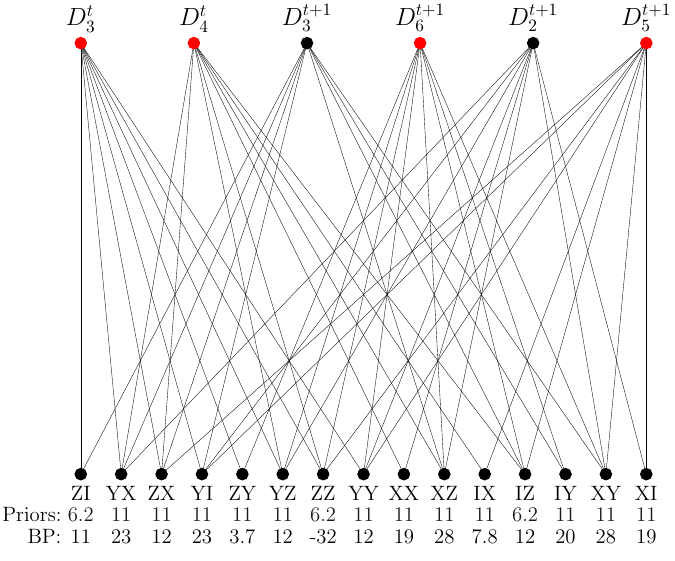}}
\caption{(a) A layout and schedule for measuring the stabilizers of an $L=3$ XY surface code.  The stabilizers are numerically labelled 1-8 at the corresponding ancilla vertex.  We show the control-not and control-$Y$ gates used to measure these stabilizers with numerical labels 0-3 indicating the time ordering of these gates.  We define a detector $D_{j}^{t}$ as the parity of stabilizer $j$ in consecutive rounds $t-1$ and $t$. (b) The circuit-level Tanner graph corresponding to the circuit in (a). We show only a small subgraph of the full Tanner graph, corresponding to the two-qubit Pauli errors that can occur after the highlighted CNOT gate in round $t$ in (a). Below each variable node, we also show the log-likelihood ratio (LLR) of its prior, as well as the LLR of the posterior probability estimate output by BP given the syndrome in which the red check nodes are flipped. The BP hard decisions here would output $ZZ$ as a correction.}
\label{fig:circuit_level_tanner_graph}
\end{figure}

Temporal boundaries arise not only during logical state preparation and measurement, but also during lattice surgery operations.
\Cref{fig:LatticeSurgery} shows two XY surface code patches being merged into a single patch, which is the first step of lattice surgery for measuring a $\overline{Y} \otimes \overline{Y}$ logical observable.
\Cref{fig:LatticeSurgery} highlights a $O(\sqrt{n})$ $Z$ error occurring just before the merge that would go undetected and cause a logical error.  The preparation of physical qubits in the $\vert + \rangle$ state in this time slice correspond to temporal boundaries in the space-time picture of \Cref{fig:SpaceTimeDiagram} where error 3 represents a similar temporal boundary error.  This is essentially the same error mechanism as afflicts logical state preparation, which can be seen from comparing errors 3 and 4 in \Cref{fig:SpaceTimeDiagram}.

Most fragile errors encountered have been constrained to boundaries, either temporal or spatial.  However, during lattice surgery a logical failure can also occur due to string-like errors propagating through the bulk as illustrated by error 5 of \Cref{fig:SpaceTimeDiagram}.  Since these errors terminate at temporal boundaries but travel through the bulk, we refer to them as temporal bulk errors. Note that a vertical error in the space-time picture corresponds to a measurement failure of a stabilizer measurement that occurs with some probability $p_m$.  If we repeat these stabilizer measurements $d_m$ times during lattice surgery, then error 5 of  \Cref{fig:SpaceTimeDiagram} represents $d_m$ consecutive measurement faults and occurs with probability $O(p_m^{d_m/2})$.  This fault results in the lattice surgery operation giving an incorrect value of the measured logical multi-qubit Pauli operator.

A standard choice is to set $d_m=\sqrt{n}$, and if $p_m$ is similar to the probability of a $Z$ error, then the probability of each such temporal bulk errors is comparable to a $O(\sqrt{n})$ $Z$ error.  However, there are more possible temporal bulk errors since there are more paths through the bulk than along the boundaries.  Of course, temporal bulk errors can be suppressed by having more rounds of stabilizer measurements during lattice surgery (e.g. setting $d_m = n$) but this results in significantly slower quantum computation.

\section{Additional numerical results for tailored surface codes}
\label{app:methods_for_numerics_tailored}

\subsection{Impact of state preparation and measurement errors}
\label{app:state_prep_measure_numerics}

\begin{figure}
\centering
\includegraphics[width=0.9\columnwidth]{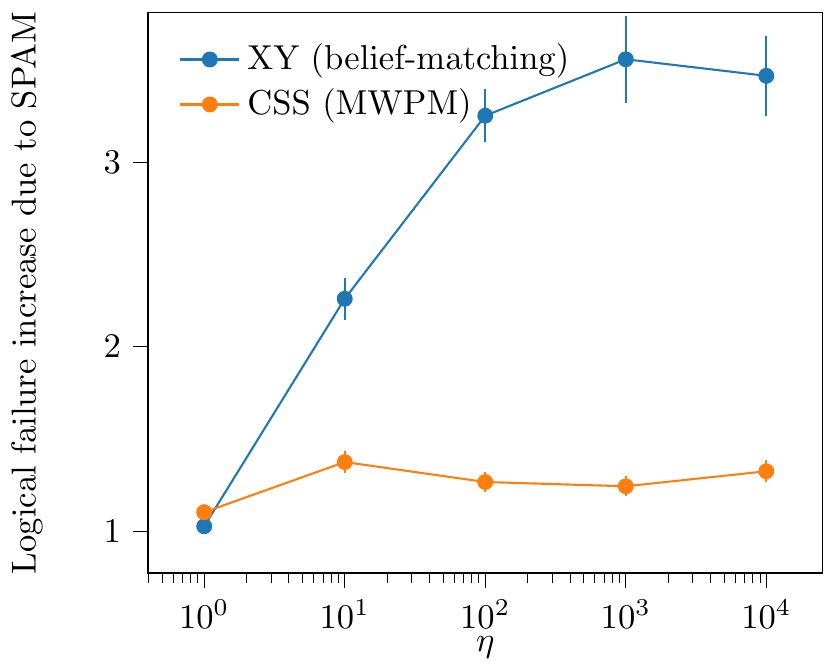}
\caption{
Effect of SPAM errors on the logical error rate of the $L=7$ XY and CSS surface codes, for $L$ rounds of noisy syndrome extraction and $p=0.015$.
The $y$ axis shows the ratio $p_{\mathrm{log}}^{\mathrm{SPAM}}/p_{\mathrm{log}}^{\mathrm{mem}}$ where $p_{\mathrm{log}}^{\mathrm{SPAM}}$ is the logical error rate including logical SPAM errors, and $p_{\mathrm{log}}^{\mathrm{mem}}$ is the logical error rate using perfect logical state preparation and measurement.
}
\label{fig:spam_vs_no_spam}
\end{figure}

In order to better understand the effect of the $O(\sqrt{n})$ $Z$-type failure mechanisms present during logical SPAM (fragile temporal boundary errors), we simulated the XY surface code using perfect SPAM, as well as noisy SPAM. 
For perfect SPAM, we use a round of perfect syndrome extraction at the beginning and end of the computation.
For noisy SPAM, we initialise data qubits in the $\ket{+}$ state before the first round, and measure data qubits in the $X$ basis at the end of the computation (with physical state preparation and measurement errors occurring at the rate given by the noise model), and all rounds of syndrome extraction circuits are noisy.
Using a $L=7$ XY surface code with $L$ rounds of noisy syndrome extraction, we then calculate the ratio $p_{\mathrm{log}}^{\mathrm{SPAM}}/p_{\mathrm{log}}^{\mathrm{mem}}$, where $p_{\mathrm{log}}^{\mathrm{SPAM}}$ is the logical error rate using noisy SPAM, and $p_{\mathrm{log}}^{\mathrm{mem}}$ is the logical error rate using perfect SPAM.
We also carry out the same analysis for the square, CSS surface code decoded using MWPM.
As shown in \Cref{fig:spam_vs_no_spam}, the ratio $p_{\mathrm{log}}^{\mathrm{SPAM}}/p_{\mathrm{log}}^{\mathrm{mem}}$ increases significantly with bias for the XY surface code, but remains small and approximately constant for the square, CSS surface code.
This is consistent with the $Z^{\otimes L}$ errors that can occur during SPAM being more probable than fragile spatial boundary errors.
Note that the ratio $p_{\mathrm{log}}^{\mathrm{SPAM}}/p_{\mathrm{log}}^{\mathrm{mem}}$ will also depend on the lattice size and the number of rounds, and we would expect the ratio to decrease as the number of rounds is increased.

\subsection{Below threshold scaling of the XY surface code}
\label{app:xy_below_threshold}

\begin{figure}
\centering
\includegraphics[width=0.9\columnwidth]{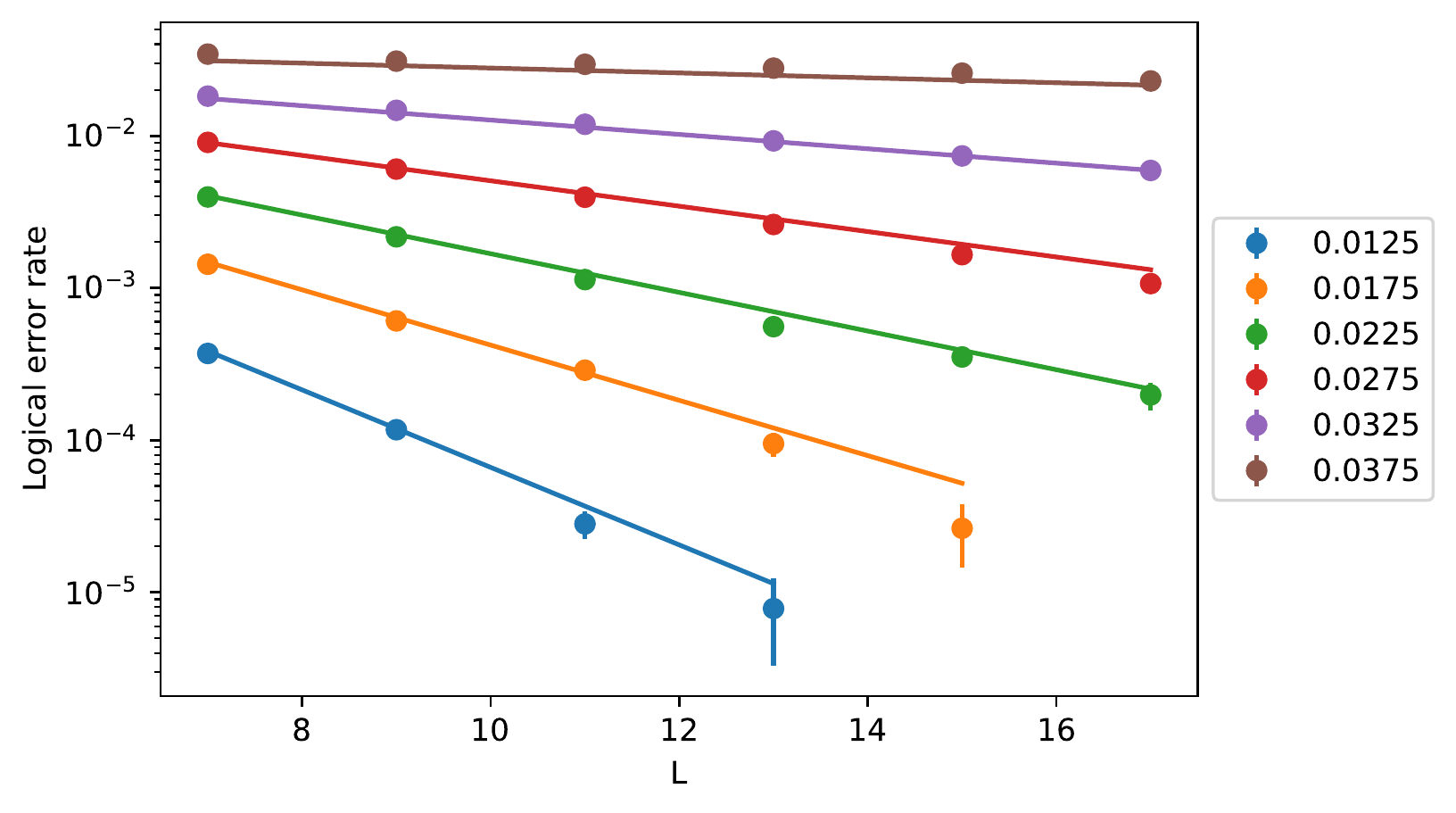}
\caption{
Logical error rate for a XY surface code with belief-matching. The legend gives the noise rate $p$ at $\eta=100$.
}
\label{fig:tailored_below_threshold}
\end{figure}

Here we present our numerical analysis of the below threshold scaling of the XY surface code.
Our results are shown in \Cref{fig:tailored_below_threshold}, where we find that our data is a good fit for an ansatz of the form 
\begin{equation}\label{eq:xy_ansatz}
    p_{\mathrm{log}}=a_{\mathrm{tailored}}(b_{\mathrm{tailored}}p)^{(\sqrt{n}+1)/2}
\end{equation}
for which we find $a_{\mathrm{tailored}}=0.0419(6)$ and $b_{\mathrm{tailored}}=24.76(7)$.
Here, $p_{\mathrm{log}}$ is the logical $Y$ error rate.
Since there is symmetry of the schedule in the bulk (a rotation and reflection of the lattice followed by an exchange of $X$ and $Y$), we expect (and have numerically verified) the logical $X$ error rate to be almost identical to the logical $Y$ error rate due to the $X/Y$ symmetry of the noise model.
The fit to this ansatz enabled us to estimate the qubit overhead required by the XY surface code to achieve a target logical error rate of $10^{-12}$.

\subsection{Below threshold scaling of the rectangular CSS surface code}

\begin{figure}
\centering
\includegraphics[width=0.9\columnwidth]{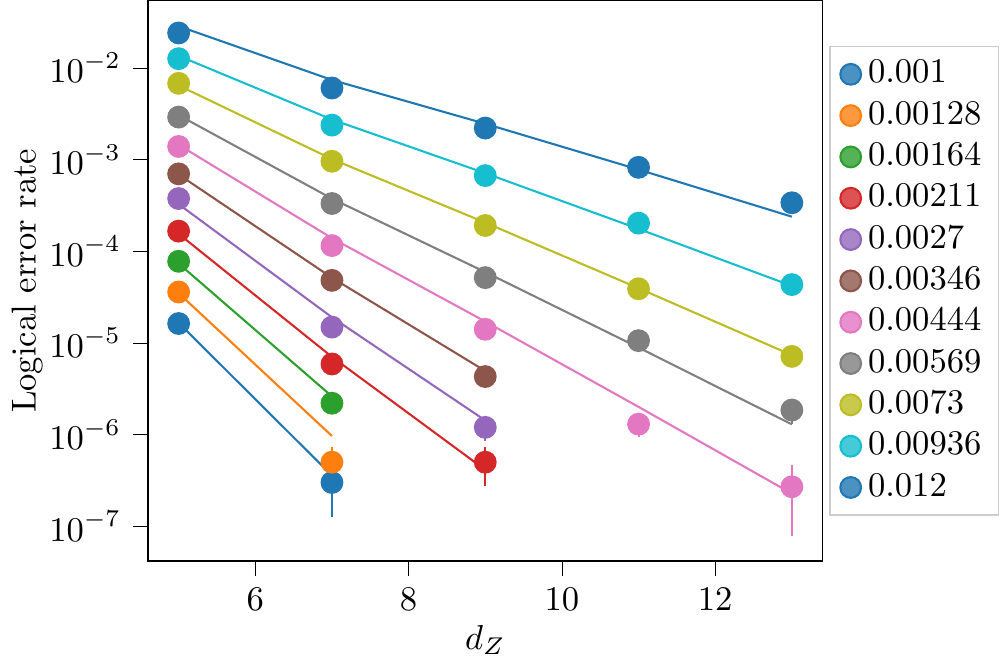}
\caption{Below threshold scaling of the rectangular, CSS surface code. Using an $X$ distance of 7 (X axis specifies $Z$ distance), and $\max(d_X,d_Z)$ rounds of stabilizer measurements. The legend gives the noise rate $p$.}
\label{fig:xz_below_threshold_performance}
\end{figure}

\begin{figure}
\centering
\includegraphics[width=0.9\columnwidth]{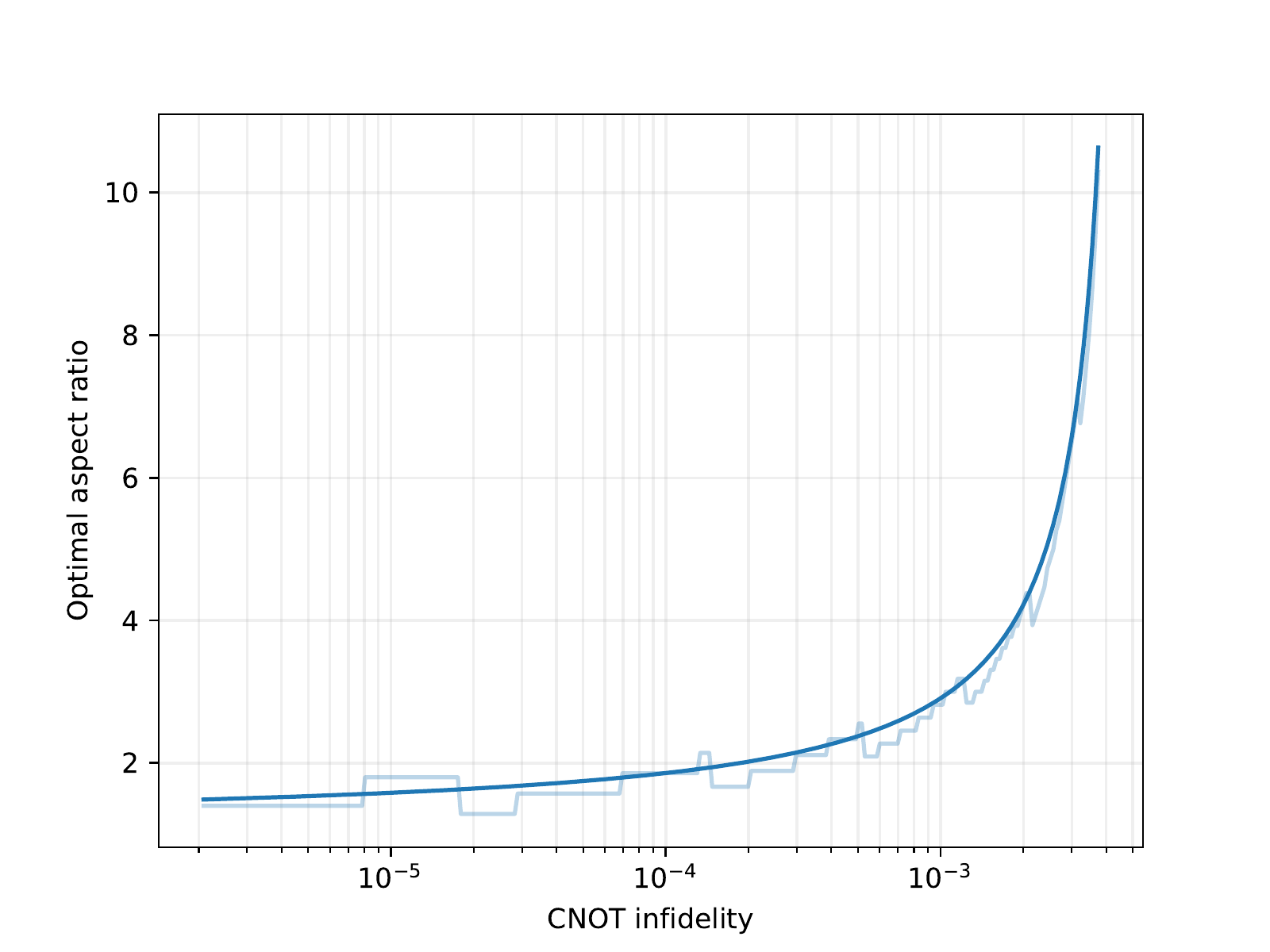}
\caption{Optimal aspect ratio for the rectangular, CSS surface code as a function of CNOT infidelity, for $\eta=100$ and a target logical failure rate of $p_{\mathrm{log}}=10^{-12}$. For the translucent line, code distances were restricted to odd integers. These aspect ratios were calculated by setting $p_{\mathrm{log}}^X=p_{\mathrm{log}}^Z=p_{\mathrm{log}}/2$ for the fitted ans\"atze in \Cref{eq:xz_ansatz_1} and \Cref{eq:xz_ansatz_2}.}
\label{fig:xz_optimal_aspect_ratios}
\end{figure}

In \Cref{fig:xz_below_threshold_performance}, we show the performance of the rectangular CSS surface code below threshold for various code distances.
This is a subset of a larger dataset that we use to fit an ansatz of the form
\begin{align}
    p_{\mathrm{log}}^X&=\frac{a_xrd_Z}{d_X^2}(b_xp)^{(d_X+1)/2}\label{eq:xz_ansatz_1}\\
    p_{\mathrm{log}}^Z&=\frac{a_zrd_X}{d_Z^2}(b_zp)^{(d_Z+1)/2}\label{eq:xz_ansatz_2}
\end{align}
where $r$ is the number of rounds of syndrome extraction, $d_X$ and $d_Z$ are the $X$ and $Z$ distances, and $p_{\mathrm{log}}^X$ and $p_{\mathrm{log}}^Z$ are the $X$ and $Z$ logical error rates, respectively.
For a bias of $\eta=100$, we found fit parameters $a_x=0.1015(9)$, $b_x=42.30(7)$, $a_z=0.0527(9)$ and $b_z=1.69(1)$.

From these ans\"atze we find the optimal aspect ratios for a target logical failure rate of $p_{\mathrm{log}}=10^{-12}$, such that the $X$ and $Z$ logical failure rates are equal.
These optimal aspect ratios are shown in \Cref{fig:xz_optimal_aspect_ratios}, and are used to estimate the qubit overhead of the CSS surface code in the main text.

\section{Mitigating fragile errors}\label{sec:mitigating_fragile_errors}

\begin{figure*}
    \centering
    \includegraphics{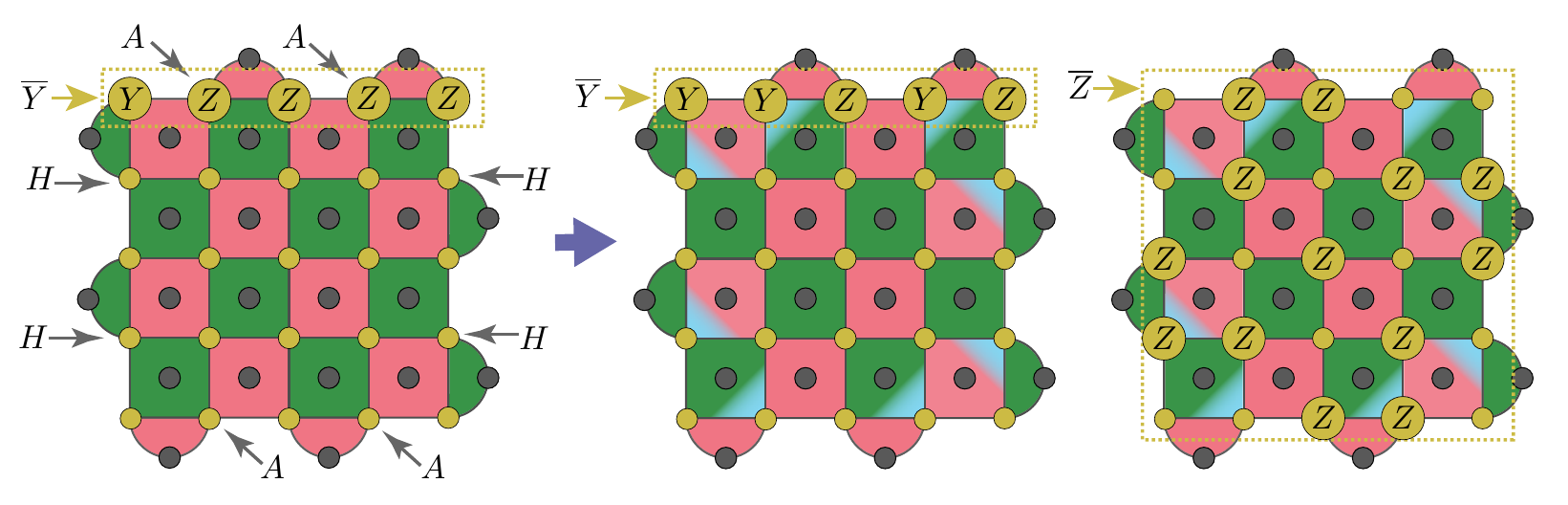}
    \caption{Deforming the boundary of the XY surface code by applying single-qubit Clifford operators $H$ and $A$, which swap $X\leftrightarrow Z$ and $Y\leftrightarrow Z$ operators, respectively.
    On the left, we show the XY surface code and a representative of a logical $\overline Y$ prior to the deformation.
    After the deformation, any logical operator on the boundary comprises at least $3$ Pauli $X$ or $Y$ operators, making the XY surface code with deformed boundary more robust.
    At the same time, a $Z$-type logical operator can be realised with fewer than $n$ Pauli $Z$ operators, as illustrated in the rightmost example.}
    \label{fig:DoublyTailored}
\end{figure*}

In this section we present a modification to the XY surface code that partially mitigates fragile spatial boundary errors at finite bias.
Recall that the XY surface code is prone to fragile spatial boundary errors composed of
a single $X$ or $Y$ error and $\sqrt{n}-1$ $Z$ errors running along the lattice boundary.
We can apply single-qubit Clifford operators along some qubits on the boundary as in \Cref{fig:DoublyTailored} so that this boundary error has $\sim \sqrt{n}/2$ $Y$ (or $X$) errors.
More specifically, we apply the Hadamard gate $H$ to one of the two qubits in the support of each $Y$ boundary stabilizer, and the $A \coloneqq HSH$ gate, where $S$ is the phase gate, to one of the two qubits in the support of each $X$ boundary stabilizer.
We will refer to this code as the XY surface code with deformed boundaries.  Note that it is not important which of the two qubits the Clifford is applied to in each boundary stabilizer, since the two choices are equivalent up to multiplication by the same boundary stabilizer. In \Cref{fig:DoublyTailored}, we illustrate how a deformed boundary requires more $X$ and $Y$ errors to realise a logical $X$ or $Y$, and therefore partially mitigates the fragility of spatial boundaries.  However, there is a tradeoff.  After deforming boundaries, a $\overline{Z}$ can be realised using fewer than $n$ $Z$ errors. As such, boundary deformation will impair performance at infinite bias when only $Z$ errors occur, while providing a performance boost at modest bias.  We promised only \textit{partial} progress, since our boundary deformation mitigates fragility of spatial boundaries, but it leaves open whether one can also mitigate against fragile temporal boundaries during SPAM operations and lattice surgery.  

\begin{figure}
    \centering
    \includegraphics[width=0.8\columnwidth]{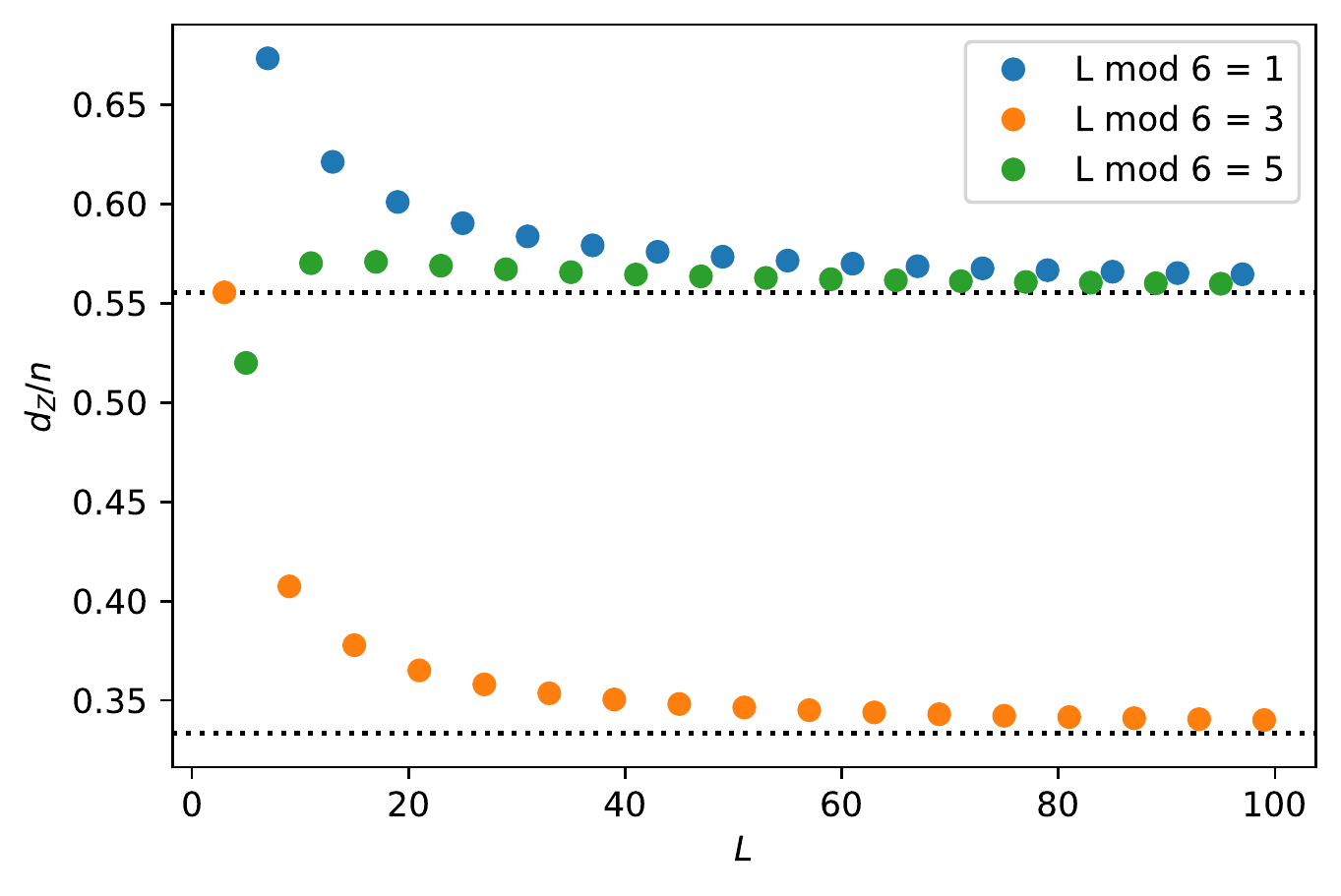}
    \caption{The $Z$ distance $d_Z$ as a fraction of the number of qubits $n=L^2$ for the XY surface code with deformed boundaries. The horizontal dotted lines are at 5/9 and 1/3.}
    \label{fig:z_type_distances_deformed_boundaries}
\end{figure}

To better quantify the effects of boundary deformation, we next consider how the weight of $Z$-type logical errors scale with the code size.
With infinite $Z$ bias, we need only consider the $X$ or $Y$ components of each stabilizer, and whether or not a stabilizer is $X$-type or $Y$-type has no impact on its syndrome.
We can therefore construct a binary linear code, where the $X$ or $Y$ component of each stabilizer corresponds to the nonzero elements of a parity check, each a row in a check matrix $\mathbf{H}$.
If a binary vector $\mathbf{v}$ is in the kernel $\ker(\mathbf{H})$ of $\mathbf{H}$, then the $Z$-type Pauli operator $\bigotimes_{i=0}^{n-1}Z^{v[i]}$ is either a $Z$-type stabilizer or $Z$-type logical operator (here $Z$-type refers to a Pauli operator in $\{I, Z\}^{\otimes n}$).
We can also easily check whether an element in $\ker(\mathbf{H})$ is a stabilizer or non-trivial logical operator by determining if it commutes with the logical $X$ and $Y$ operators of the code.
Using this approach, we computed all $Z$-type logical operators and stabilizers of the XY surface code with deformed boundaries for all odd $L<100$.
We denote by $d_Z$ the $Z$ distance.

\begin{figure}
\centering
\subfloat[\label{subfig:doubly_tailored_L15}]{
\includegraphics[width=0.3\columnwidth]{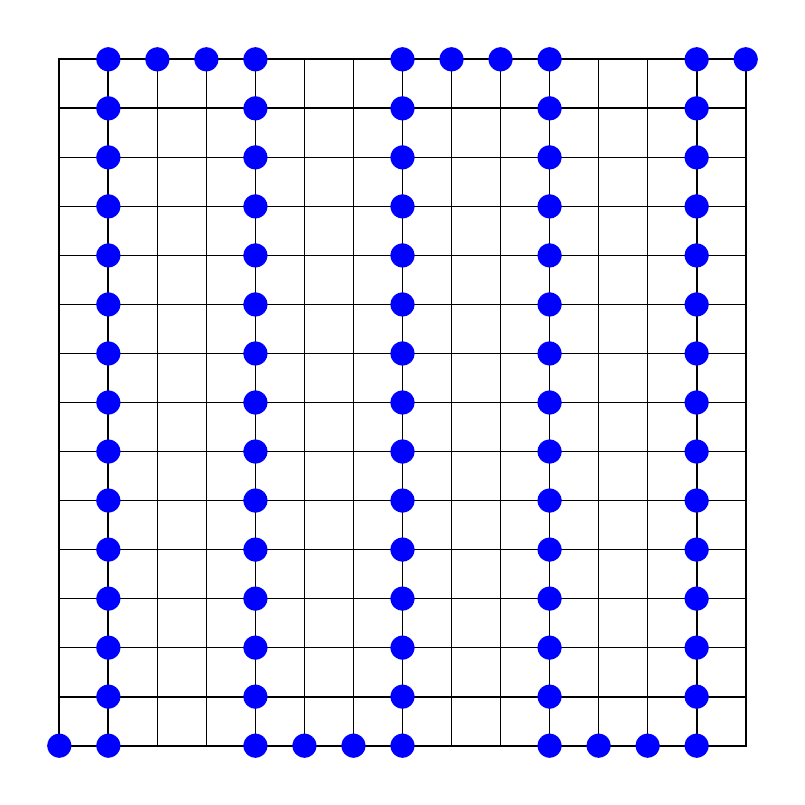}}\hfill
\subfloat[\label{subfig:doubly_tailored_L17}]{
\includegraphics[width=0.3\columnwidth]{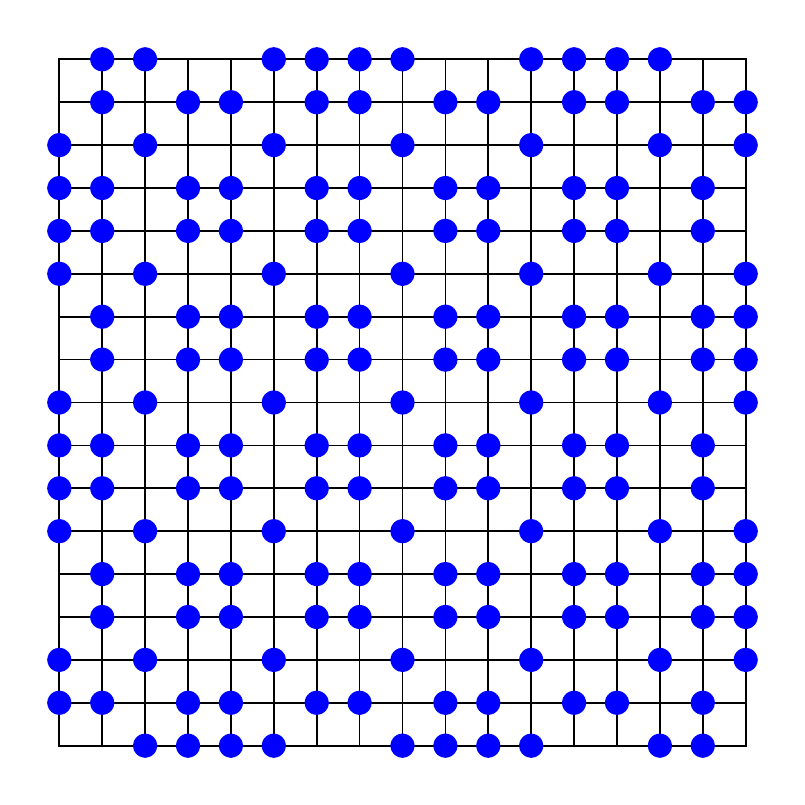}
}\hfill
\subfloat[\label{subfig:doubly_tailored_L19}]{
\includegraphics[width=0.3\columnwidth]{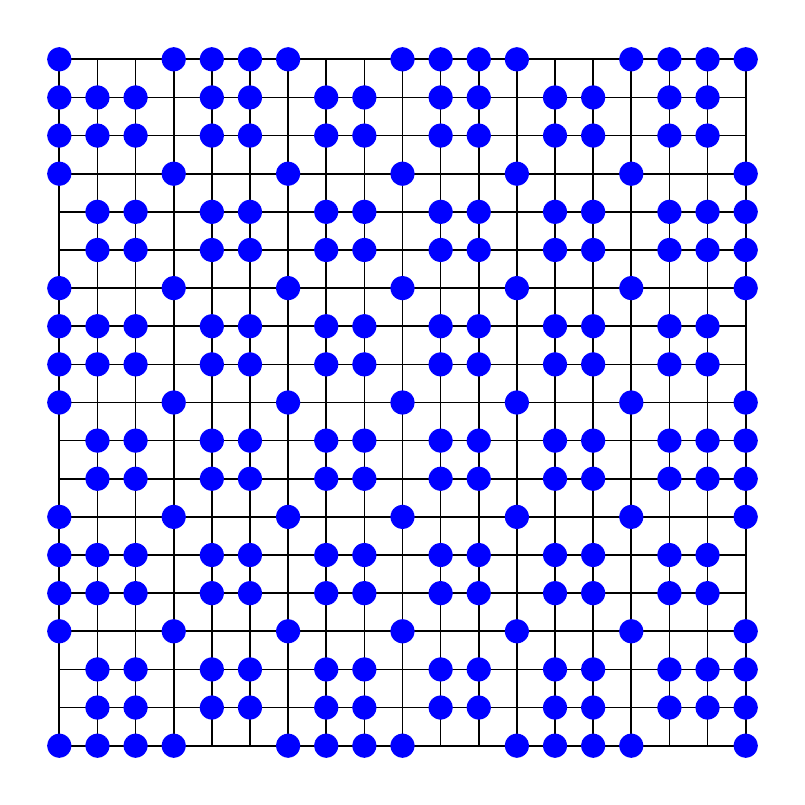}}\hfill
\caption{The $Z$-type logicals of the XY surface code with deformed boundaries for (a) $L=15$, (b) $L=17$ and (c) $L=19$. Each blue marker denotes a $Z$ operator. Stabilizers are on faces and are of the same form as in \Cref{fig:DoublyTailored}. For the codes in (b) and (c) the $Z$-type logical is unique and there are no $Z$-type stabilizers. For the code in (a) there is also a $Z$-type stabilizer which, multiplied by the $Z$ logical in (a) gives another $Z$ logical of the same form but rotated $90^\circ$.}
\label{fig:doubly_tailored_z_logicals}
\end{figure}

We found that the deformed boundaries degrade the $Z$ distance by only a constant factor, so that it still scales as $\Omega(n)$.
In \Cref{fig:z_type_distances_deformed_boundaries} we show the ratio $d_Z/n$ for all odd $L<100$.
For large $L$, $d_Z/n$ converges to $5/9$ if $L\mod 6 = 1, 5$, and converges to $1/3$ if $L\mod 6 = 3$.
This can be understood by considering the structure of the $Z$-type logical operators, as shown for codes with
$L=15,17,19$ in \Cref{fig:doubly_tailored_z_logicals}.
For $L=15$, the $Z$-type logical forms a square wave that traverses the lattice.
Considering a $3\times 3$ unit cell in the bulk of the lattice, we see that the logical operator has support on one third of the qubits.
There is an equivalent $Z$-type logical operator obtained by a $90^\circ$ rotation, related by a $Z$-type stabilizer.
There are no other $Z$-type stabilizers or logical operators for this code (the dimension of $\ker(\mathbf{H})$ is 2).
Both the $L=17$ and $L=19$ codes are tiled by a $3\times 3$ unit cell with the same structure, but with different boundaries for the two codes.
Within a $3\times 3$ unit cell of these two codes, the $Z$-type logical operator has nontrivial support on 5/9 qubits.
For both $L=17$ and $L=19$ there is only one $Z$-type logical operator (which is symmetric under $90^\circ$ rotations) and no $Z$-type stabilizers.
For all three of these codes, the structure of the $Z$-type logical operator and the code itself are periodic, both horizontally and vertically, with a period of 6.
Therefore, adding 6 columns (or rows) to the lattice leaves the structure of the $Z$-type logical operator unchanged.
As a result, the $Z$-type logical operators for $L=15$, $L=17$ and $L=19$ generalize for all $L>6$.
This structure and periodicity of the $Z$-type logical operators explains the data in \Cref{fig:z_type_distances_deformed_boundaries}, and the convergence of $d_Z/n$ to 5/9 and 1/3.
Therefore, at infinite bias, deforming the boundaries degrades performance relative to the XY surface code (reducing $d_Z$ from $n$ to $5n/9$ or $n/3$), however we expect performance to improve substantially for noise with (even very large) finite bias.

As discussed in \Cref{sec:performance_below_threshold}, we expect the logical failure rate associated with the fragile spatial boundary errors of the XY surface code to decay as $O(p^{\sqrt{n}/2+O(1)}/\sqrt{\eta})$ far below threshold.
By deforming the boundaries, we expect string-like errors along the spatial boundaries to occur with probability $O(p^{\sqrt{n}/2}\eta^{-\sqrt{n}/4})$.
On the other hand, by deforming the boundaries we now have pure $Z$-type errors occurring with probability $O(p^{cn/2})$, where here $cn=d_Z\geq n/3$ is the weight of the $Z$-type logical operator; see \Cref{fig:z_type_distances_deformed_boundaries}.
This $O(p^{cn/2})$ scaling is worse than the $O(p^{n/2})$ scaling of $Z$-type logical failures in the XY surface code (with undeformed boundaries), however for the pure $Z$-type errors to dominate we would require extremely high biases for any reasonable choice of $p$ and $n$.
Note that this analysis has only considered a few specific failure mechanisms, and a more detailed analysis of other failure mechanisms (as well as a consideration of entropic contributions to the error rate and circuit-level simulations) will be crucial to better understand and quantify the potential improvement.

\end{document}